\documentclass[conference]{IEEEtran}

\usepackage[T1]{fontenc}
\usepackage[utf8]{inputenc}
\usepackage{amsmath,amssymb,amsthm,mathtools}
\usepackage{bm}
\usepackage{enumitem}
\usepackage{hyperref}
\usepackage{booktabs}
\usepackage{microtype}
\usepackage{graphicx}
\usepackage{subcaption}
\usepackage{float}
\usepackage{xcolor}
\usepackage[section]{placeins}
\usepackage{dblfloatfix}

\usepackage[round]{natbib}
\title{Dynamic Collateral Control for Permissionless Spot Perpetual Basis Trading}

\author{
Anatoly Krestenko$^{1,2}$ Mikhail Butov$^{4}$ \\
Rostislav Berezovskiy$^{2}$, and 
Danila Bolotin$^{2,3}$ \\
\\
$^{1}$Moscow Institute of Physics and Technology \\
$^{2}$Vega Institute Foundation \\
$^{3}$Moscow State University \\
$^{4}$Independent Researcher \\
}

\newtheorem{remark}{Remark}

\begin{document}

\makeatletter
\def\@biblabel#1{#1.}   
\makeatother

\maketitle

\begin{abstract}
We study permissionless spot--perpetual basis trading in decentralized finance as a collateral control problem. The strategy holds spot inventory, hedges directional exposure with a short perpetual, and allocates capital between spot inventory and derivative margin under on-chain liquidity and execution frictions.

The paper delivers three results. First, it solves a static control problem for the collateral share and shows that the risk-constrained formulation provides a more robust operating benchmark relative to the economic optimum. In comparative calibration, the required collateral rises monotonically under volatility stress. The collateral is the lowest for BTC and increases significantly for long tail assets such as LINK and DOGE. Second, the paper derives an asymmetric dynamic extension in which the lower boundary of intervention is solvency driven, and the upper boundary is determined by a trade-off between carry-loss and the cost of rebalancing. Monte Carlo simulation shows that the lower boundary remains structurally relevant, whereas meaningful interior upper triggers survive mainly in the regimes with high carry and low costs. Third, the paper validates an execution-aware implementation with live routed execution and historical backtests. The execution layer shows that the realized wedges are significant, but become worse in the case of selling the basis. This justifies a minimum effective rebalancing size and a positive execution buffer. The historical validation shows that in the case of a fixed control rule the realized performance is predominantly explained by the funding environment.
\end{abstract}

\begin{IEEEkeywords}
spot--perpetual basis trading, decentralized finance (DeFi), perpetual futures, collateral control, liquidation risk, dynamic band control, execution costs, funding rate arbitrage
\end{IEEEkeywords}

\section{Introduction}

Basis trading is a canonical relative-value trade: the trader acquires spot inventory, sells a futures contract against it, and monetizes the wedge between spot financing and derivative carry. In crypto markets, the natural analog is the spot--perpetual basis trade, in which the carry is realized through periodic funding transfers on the perpetual contract \citep{AckererHugonnierJermann,HeManelaRossWachter}. We formulate the permissionless version of that trade as a control problem in decentralized finance (DeFi).

The architecture is distinctive in economic terms. The trading agent sources spot inventory from a decentralized exchange (DEX) and establishes the hedge through a perpetual short on a decentralized perpetual venue (PerpDEX). In concentrated-liquidity AMMs such as Uniswap~v3, the executable depth is local in the price space and therefore depends on the liquidity currently posted in the relevant range \citep{UniswapV3Whitepaper,LiaoDominance}. Perpetual trading is also available on decentralized venues, where market design ranges from virtual AMMs to oracle-based and order-book-based systems \cite{ChenMaNiePerpDEX}. Hyperliquid is particularly relevant because it combines on-chain liquidity and perpetual trading with transparent leverage, liquidation, and margin rules \cite{HyperliquidLiquidations,HyperliquidMargining}.

The economic question is whether the permissionless basis trade admits a viable allocation and intervention policy once one prices both solvency and execution. Relative to centralized desks, the permissionless implementation is less immediate because spot inventory must be sourced from finite on-chain liquidity and may incur slippage. At the same time, the decentralized stack offers two structural advantages. First, collateral management, intervention logic, and settlement can be encoded in smart contracts executed on a public state-transition system \cite{EthereumWhitepaper,YellowPaper}. Second, decentralized derivative venues list a broader cross-section of long-tail assets whose funding premia can materially exceed those of large-cap benchmark assets. The relevant trade-off is therefore capacity and immediacy on one side versus automation and access to broader carry opportunities on the other. This framing is also consistent with the broader literature on transaction costs and dynamic trading, which emphasizes that portfolio control must be derived jointly with execution frictions rather than imposed ex post \cite{AlmgrenChriss2001,GarleanuPedersen2013,ObizhaevaWang2013}.

The empirical design follows this interpretation. Binance serves as the centralized benchmark and Hyperliquid as the decentralized derivatives benchmark, while spot execution is interpreted as DEX-based and later validated with live routed execution. The static problem is the correct starting point because the first-order decision is the collateral share, not yet the timing of repeated interventions. The dynamic layer is then built around the static benchmark target: the trading agent keeps the realized collateral share inside an admissible band, with a solvency-driven lower boundary and an economically filtered upper boundary. This asymmetry is structural. Liquidation risk is immediate, whereas downward rebalancing is justified only if the carry recovered by freeing collateral dominates the implementation cost.

To keep the empirical layers comparable, the paper uses a simple unit-and-horizon convention throughout. Funding enters the static and dynamic layers through horizon-matched averages and cumulative carry terms, execution costs are reported in basis points, realized volatility is annualized from hourly returns, and short horizon liquidation control is indexed by the liquidation horizon $h_{\mathrm{liq}}$. This convention is maintained in the refreshed comparative calibration, the historical backtest calibration, and the live execution layer.

The contribution has three parts. The first is static: we derive the collateral allocation problem and show that the risk-constrained solution is a more useful operating benchmark after calibration. The second is dynamic: we derive an asymmetric band-control rule in which the lower trigger is dictated by a short horizon liquidation budget, and the upper trigger by a comparison between the carry-loss and the cost of rebalancing. The third is implementation: we validate the resulting policy with historical backtests and live execution diagnostics.

The remainder of the paper is organized as follows. Section~II reviews the relevant literature. Sections~III--VII develop the market architecture, model, static control problem, and comparative calibration. Section~VIII derives the dynamic extension. Section~IX reports the historical backtest layer. Section~X studies live execution validation. Section~XI concludes.
\section{Related Work}

The paper combines several strands of literature.

The first strand concerns the pricing and economics of perpetual futures. \cite{AckererHugonnierJermann} derive no-arbitrage valuations for perpetual futures under several contract specifications, while \cite{HeManelaRossWachter} characterize perpetual prices and bounds in the presence of funding transfers and trading frictions. These papers justify treating the spot--perpetual spread and its associated funding leg as economically meaningful carry objects. However, they do not formulate the collateral allocation problem of a permissionless implementation with long spot, short perpetual futures, and explicit intervention bands.

The second strand studies the basis trading under leverage, funding fragility, and margin stress. Evidence from the Treasury cash--futures basis shows that apparently low risk convergence trades can become fragile once leverage, financing, and margin constraints bind \cite{BarthKahnTreasuryBasis,KruttliMoninPetrasekWatugala}. The lesson for the present setting is conceptual rather than asset-specific: basis trades are jointly constrained by expected carry, collateral design, and forced unwind mechanics. Our static control problem and the lower dynamic intervention boundary are direct descendants of that insight, but they are developed for a DeFi environment in which margin is posted on a PerpDEX and spot inventory must be sourced from on-chain liquidity.

The third strand concerns dynamic trading and execution under transaction costs. \cite{AlmgrenChriss2001} derive an optimal execution frontier under temporary and permanent impact, \cite{ObizhaevaWang2013} show that order book resilience is central for optimal trading, and \cite{GarleanuPedersen2013} derive dynamic portfolio rules under return predictability and trading costs. This literature provides the general control-theoretic language for frictional intervention problems. Our contribution is to adapt that logic to a spot--perpetual basis trade in which the state variable is the collateral share rather than the raw inventory position.

The fourth strand studies execution quality and price discrepancies between centralized and decentralized crypto venues. \cite{MakarovSchoar2020} document persistent cryptocurrency arbitrage wedges across exchanges. \cite{BarbonRanaldoCryptoQuality} show that the quality of DEX markets is significantly affected by fixed blockchain costs, which penalize small trades more severely than large ones. Recent work of \cite{YuminagaChenSuiExecutionWelfare} on solver based DEX execution further shows that routed and off-chain competition architectures can improve execution welfare relative to vanilla AMM routing over economically relevant size ranges. These results are closest to the implementation layer of our model. Our contribution is to connect perpetual futures pricing, basis trade solvency control, permissionless execution constraints, and live execution diagnostics in a unified framework for spot--perpetual basis trading in DeFi.
\section{Market Architecture and Strategy Design}

A DEX is a permissionless spot trading venue in which execution is mediated by on-chain liquidity rather than by bilateral custodial intermediation. In constant function and concentrated liquidity AMM families, the key economic feature is that liquidity is state dependent and visible on-chain, so that executable spot size is constrained by instantaneous pool depth as shown in  \cite{UniswapV3Whitepaper,LiaoDominance}. This property is central for basis trading because the spot leg must be sourced from available on-chain inventory rather than from a centralized order book with off-chain market making.

A PerpDEX is a decentralized venue that lists perpetual derivatives and transfers funding between longs and shorts to anchor the derivative price to the underlying spot market. The literature documents several microstructure designs and emphasizes that decentralized perpetuals inherit both the carry logic of perpetual futures and the implementation constraints of on-chain execution, as is shown in \cite{ChenMaNiePerpDEX}. In this paper the short derivative leg is benchmarked on Hyperliquid, which provides liquid on-chain perpetuals with transparent leverage and liquidation rules presented in the documentation \cite{HyperliquidLiquidations,HyperliquidMargining}. The resulting implementation is thus a permissionless basis trade: long spot acquired through DEX liquidity, short perpetual established on a PerpDEX.

The trading agent controls a single static allocation variable,
\begin{equation}
    \alpha\in(0,1),
\end{equation}
interpreted as the share of total capital posted as derivative margin. The complement $(1-\alpha)$ is allocated to the spot leg. The economic trade-off is immediate: a smaller margin share increases capital efficiency and funding exposure, but reduces the liquidation buffer; a larger margin share improves solvency at the cost of lower carry per unit of capital.

For the execution layer, it is useful to distinguish the two economically relevant basis directions. A \emph{buy-basis} trade buys spot inventory and sells the perpetual hedge, whereas a \emph{sell-basis} trade sells spot inventory and buys back the perpetual hedge. Let $q$ denote the USD notional of the contemplated rebalancing. We write
\begin{equation}
    C^{\mathrm{inst}}_{\mathrm{buy}}(q,t),\qquad C^{\mathrm{inst}}_{\mathrm{sell}}(q,t)
\end{equation}
for the contemporaneous executable cost of the corresponding basis trade at time $t$. In implementation, these objects are constructed from the currently executable spot quote or route and the contemporaneously executable perpetual price. They therefore provide the natural bridge between the dynamic control layer, the live execution validation, and the liquidity constraints developed later in the paper.

\section{Spot--Perpetual Model}

\subsection{Price and Perpetual Dynamics}

Let $p_t$ denote the spot price of the risky asset $y$ in units of the numeraire asset $x$. We model the spot process as a geometric Brownian motion,
\begin{equation}
    dp_t = \mu p_t\,dt + \sigma p_t\,dW_t.
    \label{eq:gbm}
\end{equation}
Let $f_t$ denote the perpetual futures price. Following the standard perpetual futures literature, we use the reduced form tether
\begin{equation}
    f_t = \zeta p_t,
    \qquad
    \zeta := \frac{\kappa-\iota}{\kappa + r_y - r_x},
    \label{eq:zeta}
\end{equation}
where $\kappa$ is the premium sensitivity coefficient, $\iota$ is the interest component, and $r_x,r_y$ are the convenience yield or financing terms in the two assets \cite{AckererHugonnierJermann,HeManelaRossWachter}.

For a futures position of size $H\in\mathbb{R}$ and an initial margin $M_0$, the marked-to-market derivative account is
\begin{equation}
    V_t^{F}
    :=
    M_0 + H(f_t-f_0)
    - H\int_0^t \big(\iota p_r + \kappa(f_r-p_r)\big)\,dr.
    \label{eq:futures-value}
\end{equation}
Throughout, $H<0$ corresponds to a short perpetual position.

\subsection{Portfolio Construction}

Let $D>0$ denote the total initial capital. The margin account is initialized at
\begin{equation}
    M_0 = \alpha D,
\end{equation}
and the remaining capital is deployed into the spot leg, so that the spot inventory is
\begin{equation}
    Q = \frac{(1-\alpha)D}{p_0}.
    \label{eq:Q}
\end{equation}
We impose local delta neutrality at inception by choosing the futures position to offset the spot delta,
\begin{equation}
    Q + \zeta H = 0,
\end{equation}
which yields the hedge size
\begin{equation}
    H = -\frac{Q}{\zeta} = -\frac{(1-\alpha)D}{\zeta p_0}.
    \label{eq:H}
\end{equation}
The full spot--perpetual portfolio value is then
\begin{equation}
    V_t = Qp_t + V_t^F.
    \label{eq:portfolio}
\end{equation}
With \eqref{eq:H}, the instantaneous directional exposure is neutralized by construction and the portfolio is locally delta-neutral with respect to the spot price. Writing the portfolio value as
\begin{equation}
    V_t = Q p_t + M_t + H f_t,
\end{equation}
and using the local tether approximation $f_t = \zeta p_t$, the instantaneous spot delta is
\begin{equation}
    \frac{\partial V_t}{\partial p_t}
    =
    Q + H \frac{\partial f_t}{\partial p_t}
    =
    Q + \zeta H.
\end{equation}
Hence, the portfolio is delta-neutral whenever
\begin{equation}
    H = -\frac{Q}{\zeta},
\end{equation}
which is the hedge ratio imposed throughout the paper and the residual economic source of return is the funding-carry component.

\section{Liquidation Barrier and Static Control}

\subsection{Liquidation Rule}

Under the reduced form liquidation condition,
\begin{equation}
    M_0 + H(f_t-f_0) \le \theta_F |H| f_t,
    \label{eq:liqrule}
\end{equation}
where $\theta_F\in(0,1)$ is the maintenance margin fraction, substitution of \eqref{eq:H} and $f_t=\zeta p_t$ yields the barrier condition
\begin{equation}
    \frac{p_t}{p_0} \ge \frac{1}{(1+\theta_F)(1-\alpha)}.
    \label{eq:barrier}
\end{equation}
Define the barrier parameter
\begin{equation}
    z_{\theta_F}(\alpha) := \frac{1}{(1+\theta_F)(1-\alpha)}.
    \label{eq:ztheta}
\end{equation}
Liquidation is thus a first-passage event of the GBM through $z_{\theta_F}(\alpha)$.

Let $h>0$ denote the review horizon. The associated liquidation probability is
\begin{align}
    \Pi_{\mathrm{liq}}(\alpha;h)
    &=
    \frac{1}{2}\operatorname{Erfc}\!\left(
    \frac{\ln z_{\theta_F}(\alpha)}{\sigma\sqrt{2h}}
    -
    \frac{\nu\sigma\sqrt h}{\sqrt 2}
    \right)
    \notag\\
    &\quad+
    \frac{1}{2}z_{\theta_F}(\alpha)^{2\nu}
    \operatorname{Erfc}\!\left(
    \frac{\ln z_{\theta_F}(\alpha)}{\sigma\sqrt{2h}}
    +
    \frac{\nu\sigma\sqrt h}{\sqrt 2}
    \right),
    \label{eq:pi-liq}
\end{align}
where $\nu = \mu/\sigma^2 - 1/2$ as in \cite{BorodinSalminen}.

\begin{remark}[$\zeta$ cancels out of the liquidation condition]
Substituting the delta-neutral hedge $H = \frac{-(1-\alpha)D}{\zeta p_0}$ and the tether $f_t = \zeta p_t$ into the maintenance condition \eqref{eq:liqrule} yields, on the left-hand side,
\begin{align*}
M_0 + H(f_t - f_0)
&= \alpha D - \frac{(1-\alpha)D}{\zeta p_0}\,\zeta(p_t - p_0) \\
&= \alpha D - \frac{(1-\alpha)D}{p_0}(p_t - p_0),
\end{align*}
and, on the right-hand side,
\[
\theta_F |H| f_t
= \theta_F \cdot \frac{(1-\alpha)D}{\zeta p_0}\cdot \zeta p_t
= \theta_F \cdot \frac{(1-\alpha)D}{p_0}\,p_t.
\]
The tether coefficient $\zeta$ cancels in both expressions. The liquidation condition therefore depends only on the spot ratio $p_t/p_0$, the collateral share $\alpha$, and the maintenance fraction $\theta_F$, and is invariant to the perpetual-spot premium itself. This allows first-passage analysis to be carried out directly on the GBM driving the spot price, without modeling $f_t$ as a separate stochastic process.
\end{remark}

\subsection{From Maximum Leverage to the Maintenance Fraction}

For numerical calibration, we require a reduced form mapping from venue leverage limits to $\theta_F$. Let $L_{\max}$ denote the maximum leverage reported by the venue for the asset. The corresponding maximum initial-margin fraction is
\begin{equation}
    \mathrm{IM}_{\max} = \frac{1}{L_{\max}}.
\end{equation}
If maintenance is modeled as a fraction $c$ of the maximum initial margin, then
\begin{equation}
    \theta_F = c\,\mathrm{IM}_{\max} = \frac{c}{L_{\max}}.
    \label{eq:theta-general}
\end{equation}
Hyperliquid states that the maintenance margin fraction is half of the maximum initial margin fraction, \cite{HyperliquidLiquidations,HyperliquidContractSpecs}. Accordingly,
\begin{equation}
    \theta_F^{\mathrm{HL}} = \frac{1}{2L_{\max}}.
    \label{eq:theta-hl}
\end{equation}
For Binance, the exact maintenance is tiered by notional brackets and user configuration. To obtain an account independent static baseline, we use the conservative approximation
\begin{equation}
    \theta_F^{\mathrm{BN}} = \frac{1}{L_{\max}}.
    \label{eq:theta-binance}
\end{equation}
This calibration should be interpreted as a reduced form venue rule rather than as a reconstruction of account level bracket mechanics.

\subsection{Funding Calibration and Horizon Matching}

Let $\widetilde{\kappa}_h$ denote the expected cumulative funding carry over the horizon $h$. We estimate it by a rolling horizon mean: for a one day decision horizon, funding observations are first summed over rolling one day windows and then averaged across the calibration sample. This is preferable to a simple event-level mean because the control problem is posed on the review horizon rather than on a single funding timestamp. Under stable funding conditions, the estimator is close to a scaled simple mean, but it is conceptually preferable because it directly targets the one day cashflow object that enters the static optimization.

The choice is also consistent with the empirical structure of the funding rates. In Binance, the interest component is fixed at $0.03\%$ per day by default, which corresponds to $0.01\%$ per eight-hour funding interval and approximately $10.95\%$ annualized simple carry before premium adjustments, as shown in the documentation \cite{BinanceFundingFAQ}. Hyperliquid exhibits an analogous anchoring mechanism, although the higher frequency hourly rule produces a visibly wider dispersion in raw observations. In both venues, the distribution of raw funding prints retains episodic tail deviations, but the distribution of one day cumulative funding remains positively centered and becomes progressively more stable as the calibration window increases. This makes rolling one day mean a natural static baseline: it preserves the horizon payoff relevant for the control problem while smoothing timestamp level noise.

Table~\ref{tab:fund-summary} reports a comparison of the compact funding window for BTC in both venues and between the 90-day, 180-day, and 360-day calibration windows. The key point is not that all mass concentrates exactly at the venue baseline; rather, longer windows stabilize the central carry estimate, increase the share of positive one day carry realizations, and attenuate the influence of adverse short-lived episodes. In this sense, the 180-day window is best interpreted as a compromise between recency and stability, while the 360-day window serves as a deliberately smoother robustness specification. The corresponding raw funding histograms are reported in Appendix~B.

\subsection{Static Control Problems}

The trading agent faces two natural static formulations.

\paragraph*{Variant 1: Economic optimum.}
The economic objective trades off the expected funding carry against the expected liquidation loss: $LGD$ denotes the loss given default, interpreted here as a reduced form loss incurred upon liquidation.
\begin{equation}
    \alpha_{\mathrm{econ}}^\star
    \in
    \arg\max_{\alpha\in(0,1)}
    \left\{
        (1-\alpha)D\widetilde{\kappa}_h
        -
        LGD\cdot\Pi_{\mathrm{liq}}(\alpha;h)
    \right\}.
    \label{eq:v1}
\end{equation}
If the optimum is interior, then it satisfies
\begin{equation}
    -D\widetilde{\kappa}_h
    -
    LGD\cdot
    \frac{\partial}{\partial \alpha}
    \Pi_{\mathrm{liq}}(\alpha;h)
    \Big|_{\alpha=\alpha_{\mathrm{econ}}^\star}
    =0.
    \label{eq:foc}
\end{equation}

\paragraph*{Variant 2: Risk-constrained optimum.}
The operational formulation imposes a liquidation-probability budget,
\begin{equation}
    \alpha_\varepsilon^\star
    \in
    \arg\max_{\alpha\in(0,1)}
    (1-\alpha)D\widetilde{\kappa}_h
    \quad\text{s.t.}\quad
    \Pi_{\mathrm{liq}}(\alpha;h)\le \varepsilon.
    \label{eq:v2}
\end{equation}
Whenever $\widetilde{\kappa}_h>0$, the objective decreases in $\alpha$, and the probability of liquidation also decreases in $\alpha$. Hence,
\begin{equation}
    \alpha_\varepsilon^\star
    =
    \inf\{\alpha\in(0,1):\Pi_{\mathrm{liq}}(\alpha;h)\le \varepsilon\}.
    \label{eq:v2-sol}
\end{equation}
If $\Pi_{\mathrm{liq}}$ is strictly decreasing, the optimum is equivalently characterized by
\begin{equation}
    \Pi_{\mathrm{liq}}(\alpha_\varepsilon^\star;h)=\varepsilon.
\end{equation}

\begin{remark}
Both static formulations are well posed and admit at least one optimizer.

For Variant~1, the objective
\[
J_{\mathrm{econ}}(\alpha)
:=
(1-\alpha)D\kappa_e h
-
LGD \cdot \Pi_{\mathrm{liq}}(\alpha;h)
\]
is continuous in $\alpha$ on the admissible control set. Hence, on a compact domain such as $(0,1)$ with continuous extension to the boundary), the existence of a maximizer follows from the Weierstrass theorem.

For Variant~2, the feasible set
\[
\mathcal{A}_\varepsilon
:=
\{\alpha \in (0,1) : \Pi_{\mathrm{liq}}(\alpha;h) \le \varepsilon\}
\]
is closed by the continuity of $\Pi_{\mathrm{liq}}(\alpha;h)$. It is also nonempty whenever $\varepsilon>0$ is such that the constraint can be met; in particular, since the liquidation barrier diverges as $\alpha \uparrow 1$, one has $\Pi_{\mathrm{liq}}(\alpha;h)\to 0$, so sufficiently large $\alpha$ are feasible. Therefore, there is an optimizer. Moreover, if $\kappa_e h>0$, the objective $(1-\alpha)D\kappa_e h$ is strictly decreasing in $\alpha$, so the optimum is attained at the minimum feasible collateral share,
\[
\alpha_\varepsilon^\star
=
\inf\{\alpha \in (0,1) : \Pi_{\mathrm{liq}}(\alpha;h)\le \varepsilon\}.
\]
If, in addition, $\Pi_{\mathrm{liq}}(\alpha;h)$ is strictly decreasing, then this optimizer is unique.
\end{remark}

\begin{remark}
Neither formulation admits an explicit closed form solution for $\alpha^\star$. In Variant~1, the first order condition \eqref{eq:foc} is transcendental because the control enters the logarithmic, power, and complementary error function terms through \eqref{eq:pi-liq}. In Variant~2, the same difficulty appears in the inversion of the probability of hitting the barrier.
\end{remark}

\section{Calibration Protocol}

The baseline numerical design is intentionally conservative and transparent. We use a one day review horizon, hourly realized volatility computed from mark price returns, and lookback windows of 30, 90, 180, and 360 days. The baseline funding estimator is the one day rolling-horizon mean described above. Volatility stress is introduced multiplicatively through the $1.0\times$, $1.5\times$, and $2.0\times$ regimes. The benchmark assets are BTC and ETH as large-cap baselines, together with LINK and DOGE as long-tail, higher-beta assets.

The one day horizon is a natural static baseline: it is short enough to represent a daily collateral review rule, yet long enough for both cumulative funding carry and first passage liquidation risk to be economically nontrivial. The one hour volatility estimator is similarly a compromise between high frequency realism and microstructure robustness. It captures intraday path risk, which matters for liquidation, without inheriting the full noise of minute level returns.

Our benchmark calibration window is 180 days. The diagnostic evidence in Table~\ref{tab:fund-summary} provides the supporting diagnostic evidence, the detailed empirical interpretation is deferred to Section VII.A

Unless stated otherwise, the main comparative tables and dynamic control figures in Sections~VI--VII use a refreshed article calibration based on the most recent 2025--2026 sample. This refreshed calibration is used only to report current benchmark targets and comparative venue results. The paper then uses two additional empirical layers with different roles. The historical backtest calibration is frozen on Binance 2023--2024 market data but applies Hyperliquid's target margin architecture because at the original design stage Hyperliquid itself had not yet accumulated a sufficiently long history for stable return and funding calibration. Live execution calibration is based on the realized deployment sample from 2025-04-01 to 2025-12-01 and is used only for execution-aware implementation parameters such as minimum trade size, side-specific cost budgets, and execution buffer. This separation follows the actual research chronology and prevents the validation layers from inheriting the final refreshed descriptive calibration.

\section{Simulation Results}

\subsection{Funding Window Diagnostics}

The funding window diagnostics support two methodological choices used in the main simulations. First, the rolling one day mean is a defensible carry estimator because the one day cumulative funding distribution is persistently positive on both venues even when raw funding prints exhibit nontrivial tails. Second, the 180-day lookback is a reasonable benchmark window because it substantially stabilizes the central carry estimate relative to 90 days without requiring the full degree of smoothing implied by 360 days. Table~\ref{tab:fund-summary} reports the comparison of the funding window used in the calibration argument. Binance benefits materially from moving from 90 to 180 days, chiefly through a higher positive-share and a less adverse left tail. Hyperliquid exhibits an even stronger stabilization pattern, but by 180 days its one day funding distribution is already strongly positively centered. Similarly, the 180-day window is retained as the benchmark specification, with 360 days reported as a robustness check. Appendix~B complements the table with raw funding histograms for both venues.

\begin{table}[t]
\centering
\caption{Binance BTC: one day cumulative funding APY by lookback window.}
\label{tab:fund-binance-btc}
\begin{tabular}{lcccccc}
\toprule
Window & Mean & Median & Std. & $q_{0.05}$ & $q_{0.95}$ & Share + \\
\midrule
90d  & 1.33\% & 1.67\% & 4.29\% & $-5.79\%$ & 7.26\% & 64.2\% \\
180d & 3.01\% & 3.30\% & 4.35\% & $-4.32\%$ & 9.77\% & 76.0\% \\
360d & 4.17\% & 4.35\% & 4.23\% & $-3.38\%$ & 10.95\% & 84.5\% \\
\bottomrule
\end{tabular}
\end{table}

\begin{table}[t]
\centering
\caption{Hyperliquid BTC: one day cumulative funding APY by lookback window.}
\label{tab:fund-hl-btc}
\begin{tabular}{lcccccc}
\toprule
Window & Mean & Median & Std. & $q_{0.05}$ & $q_{0.95}$ & + share \\
\midrule
90d  & 3.84\% & 5.32\% & 6.84\% & $-9.55\%$ & 10.95\% & 75.1\% \\
180d & 6.22\% & 8.13\% & 6.21\% & $-5.65\%$ & 11.85\% & 84.7\% \\
360d & 9.13\% & 10.07\% & 8.47\% & $-3.30\%$ & 23.22\% & 90.3\% \\
\bottomrule
\end{tabular}
\end{table}

\begin{table}[t]
\centering
\caption{Compact funding window comparison for the benchmark asset BTC. The 180-day window materially improves carry stability relative to 90 days on both venues, while 360 days provides an additional smoothing robustness check.}
\label{tab:fund-summary}
\begin{tabular}{llcccc}
\toprule
Venue & Window & Mean & Median & $q_{0.05}$ & + share \\
\midrule
Binance      & 90d  & 1.33\% & 1.67\% & $-5.79\%$ & 64.2\% \\
Binance      & 180d & 3.01\% & 3.30\% & $-4.32\%$ & 76.0\% \\
Hyperliquid  & 90d  & 3.84\% & 5.32\% & $-9.55\%$ & 75.1\% \\
Hyperliquid  & 180d & 6.22\% & 8.13\% & $-5.65\%$ & 84.7\% \\
\bottomrule
\end{tabular}
\end{table}

\subsection{Benchmark Slice}

Table~\ref{tab:main} reports the comparative benchmark slice used throughout the discussion: Variant~2 with $\varepsilon=0.001$, volatility stress $1.5\times$, and 180-day lookback. This slice is the preferred operational baseline because it directly maps a liquidation-risk budget into a collateral rule.

\begin{table}[t]
\centering
\caption{Benchmark comparative slice: Variant~2 with $\varepsilon=0.001$, $1.5\times$ volatility stress, and 180-day lookback.}
\label{tab:main}
\begin{tabular}{llcccc}
\toprule
Venue & Asset & $\alpha^\star$ & Lev. & $\widetilde{\kappa}_h$ & $\theta_F$ \\
\midrule
Binance & BTC  & 0.123 & 7.14x & 0.82 bps & 0.0080 \\
Binance & ETH  & 0.167 & 4.99x & 0.61 bps & 0.0100 \\
Binance & LINK & 0.200 & 3.99x & 1.09 bps & 0.0200 \\
Binance & DOGE & 0.201 & 3.97x & 0.31 bps & 0.0200 \\
Hyperliquid & BTC  & 0.127 & 6.89x & 1.70 bps & 0.0125 \\
Hyperliquid & ETH  & 0.176 & 4.69x & 1.69 bps & 0.0200 \\
Hyperliquid & LINK & 0.223 & 3.49x & 2.65 bps & 0.0500 \\
Hyperliquid & DOGE & 0.224 & 3.47x & 0.61 bps & 0.0500 \\
\bottomrule
\end{tabular}
\end{table}

The benchmark slice supports three immediate conclusions. First, the cross-sectional ranking is stable across venues:
\[
\text{BTC} < \text{ETH} < \text{LINK} \approx \text{DOGE}.
\]
Second, in the refreshed comparative calibration, Hyperliquid systematically produces higher benchmark collateral shares than Binance for every asset studied. Third, the resulting reduced form venue gap is modest for BTC but widens materially for long-tail assets, precisely where the broader decentralized opportunity set is economically most relevant. Part of this venue gap is analytically mechanical. In both the risk-constrained static problem and the lower-bound dynamic problem, liquidation is defined by the barrier condition
\[
\alpha_t \le \frac{\theta_F}{1+\theta_F}.
\]
Holding the return law, the liquidation horizon, and the admissible liquidation probability fixed, a higher maintenance fraction shifts the liquidation barrier upward and therefore tightens the admissible solvency region. Hence,
\begin{equation}
    \theta_F \uparrow \quad \Longrightarrow \quad \alpha_{\varepsilon}^{\star} \uparrow
    \qquad \text{and} \qquad
    \theta_F \uparrow \quad \Longrightarrow \quad \alpha_L \uparrow.
\end{equation}
Accordingly, part of the Binance--Hyperliquid difference is not merely empirical but is directly implied by the target margin architecture itself.

\begin{figure}[t]
    \centering
    \includegraphics[width=\columnwidth]{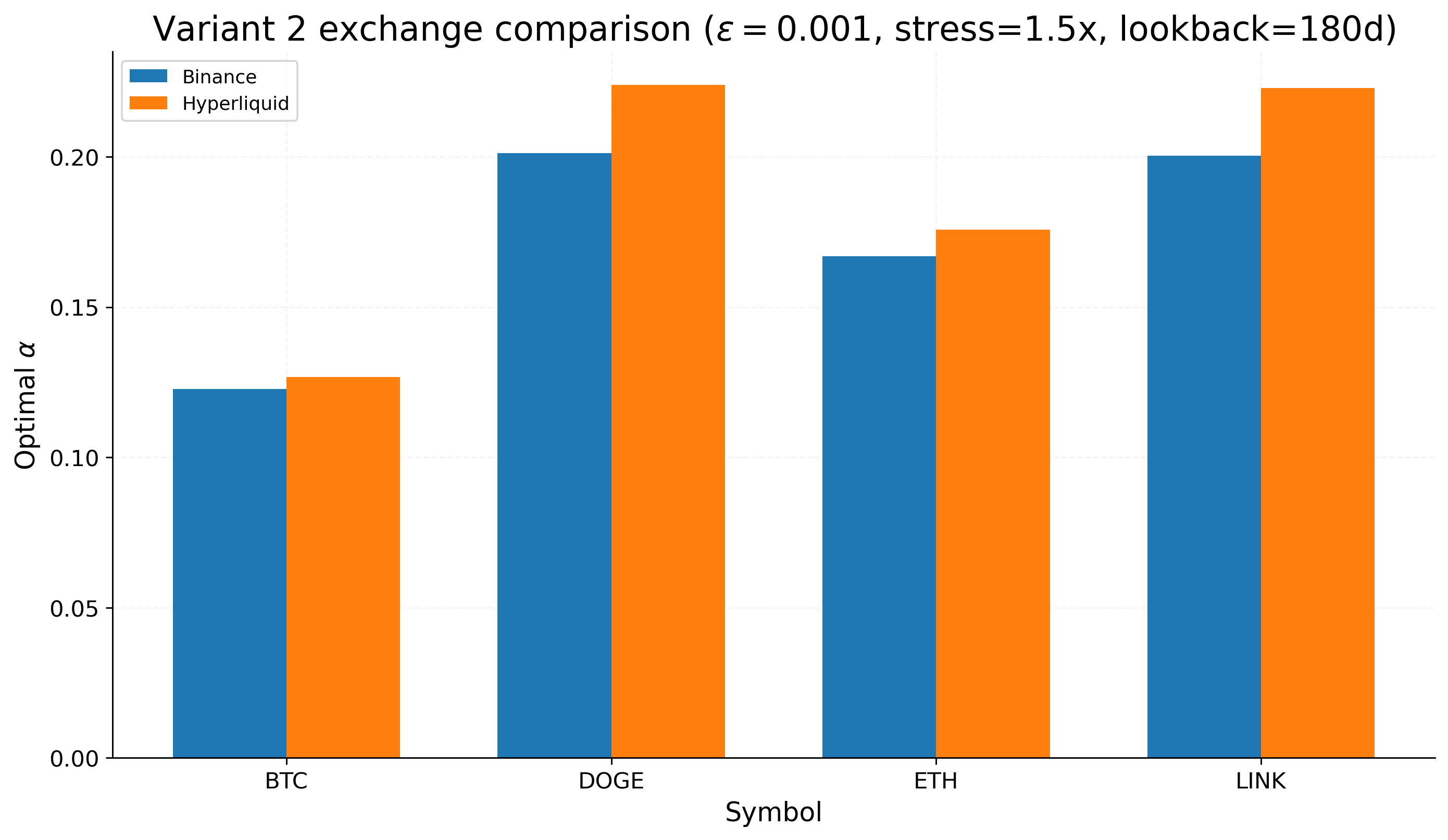}
    \caption{Direct venue comparison for Variant~2 at the benchmark slice: $\varepsilon=0.001$, $1.5\times$ volatility stress, and 180-day lookback. In the refreshed comparative calibration, the decentralized benchmark produces systematically higher benchmark collateral shares than the centralized benchmark.}
    \label{fig:bar}
\end{figure}

Figure~\ref{fig:bar} makes the venue comparison explicit: for each asset, the decentralized benchmark produces a larger benchmark margin share than the centralized benchmark. This gap should be interpreted as a reduced form comparative venue gap, reflecting the joint effect of venue specific margin architecture, realized volatility, and funding conditions, rather than as a cleanly isolated causal effect of venue design alone that the observed difference can be written as follows: 

\begin{equation}
\Delta\alpha^{\star} \approx \Delta_{\mathrm{margin}}+\Delta_{\mathrm{vol}}+\Delta_{\mathrm{funding}}
\end{equation}

\begin{remark}
The decomposition (28) is reduced-form. The tether coefficient $\zeta$ cancels in the liquidation condition, so the perpetual price enters the static problem only through its deterministic link to $p_t$. Since realized volatility is estimated from a common spot mark price series, the volatility input is venue-agnostic and $\Delta_{\mathrm{vol}} \equiv 0$ in our calibration. Cross-venue differences in $\alpha^\star$ are therefore captured by exactly two channels: the maintenance fraction $\theta_F$ ($\Delta_{\mathrm{margin}}$) and the funding stream $\widetilde{\kappa}_h$ ($\Delta_{\mathrm{funding}}$).
\end{remark}

\subsection{Comparative Sensitivity Evidence}

\begin{figure}[htbp]
    \centering
    \begin{subfigure}[t]{0.47\linewidth}
        \centering
        \includegraphics[width=\linewidth,height=0.13\textheight,keepaspectratio]{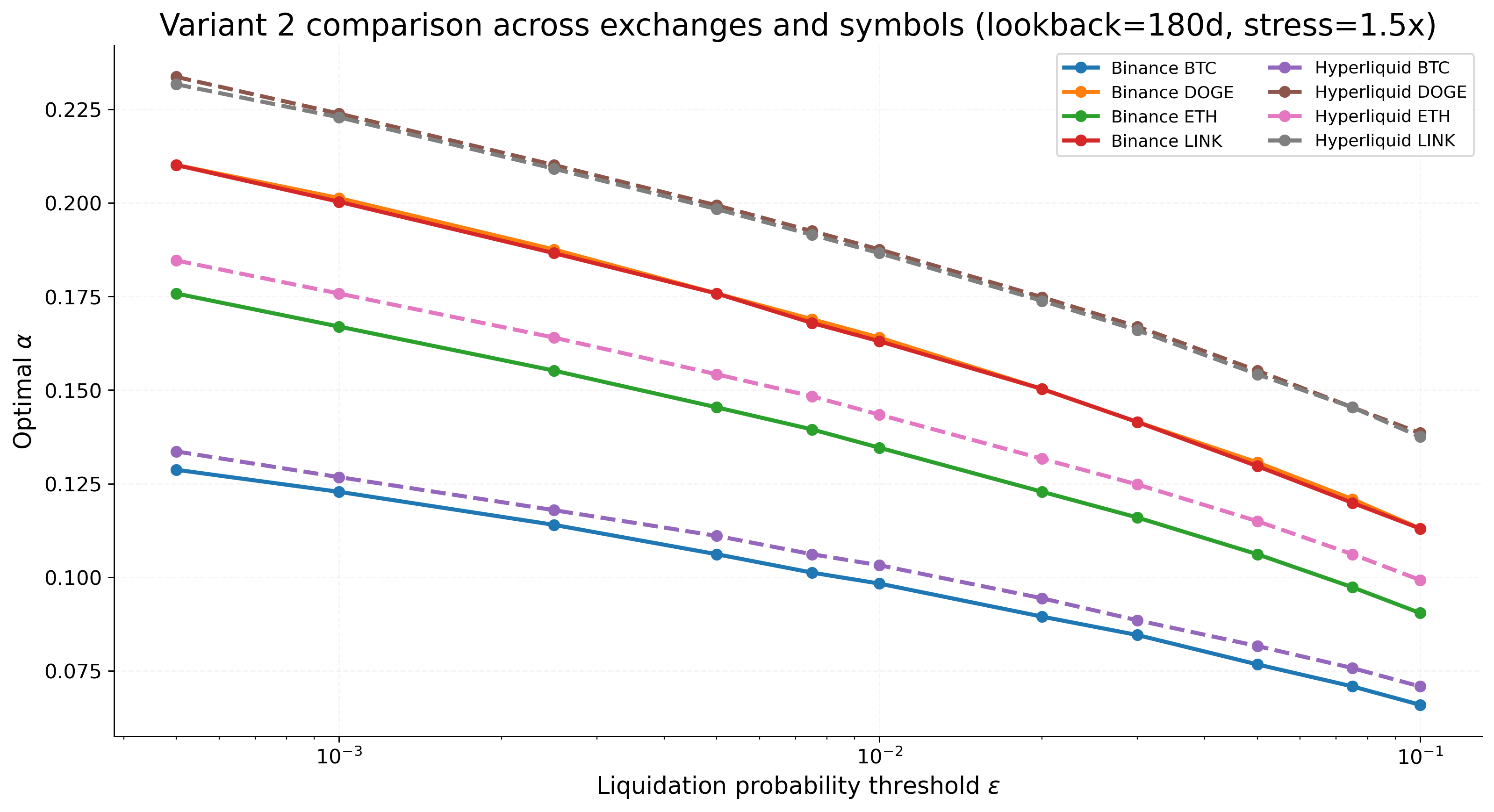}
        \caption{Variant~2 across exchanges and symbols.}
    \end{subfigure}\hfill
    \begin{subfigure}[t]{0.47\linewidth}
        \centering
        \includegraphics[width=\linewidth,height=0.13\textheight,keepaspectratio]{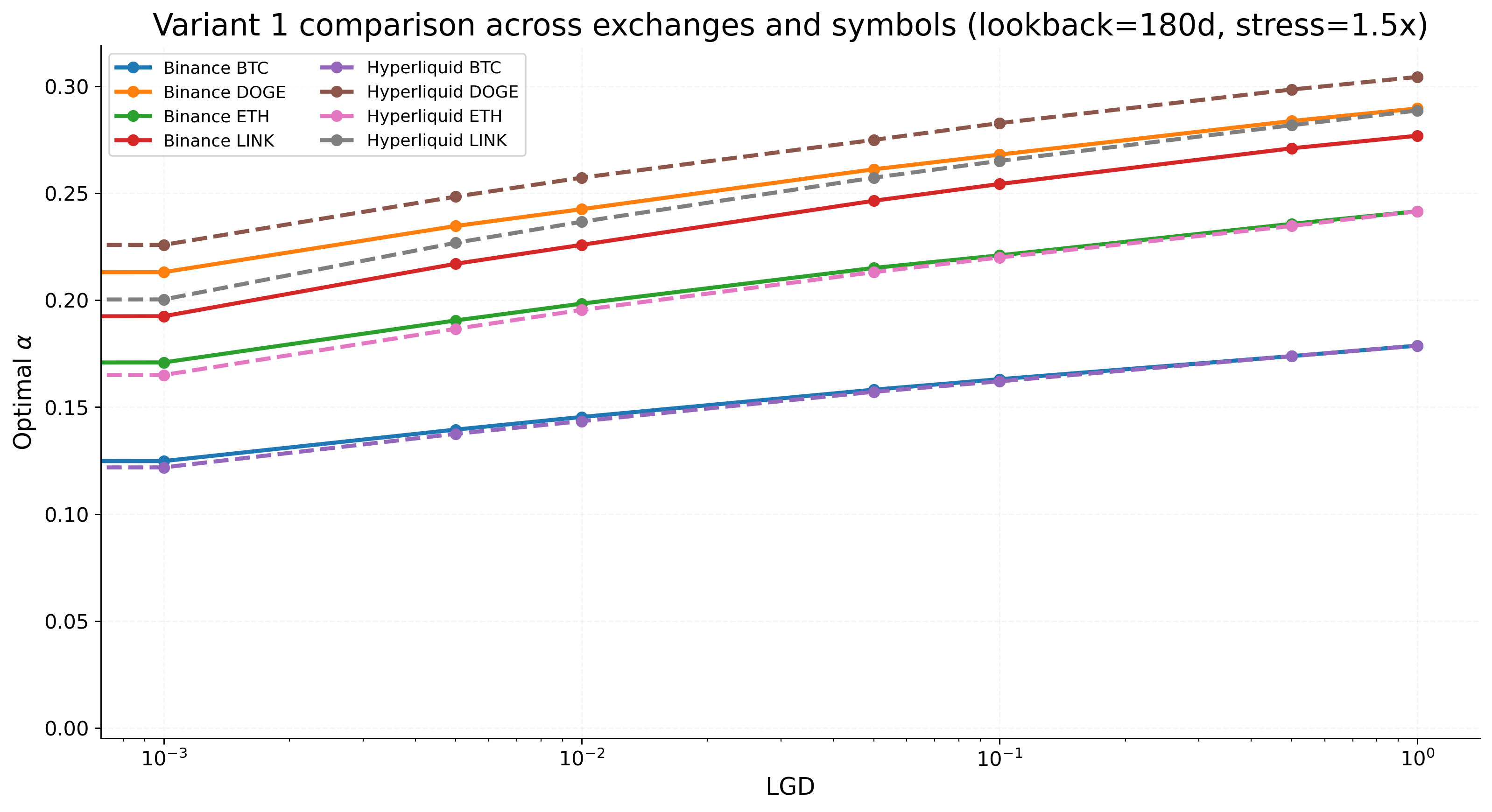}
        \caption{Variant~1 across exchanges and symbols.}
    \end{subfigure}

    \vspace{0.25em}

    \begin{subfigure}[t]{0.47\linewidth}
        \centering
        \includegraphics[width=\linewidth,height=0.13\textheight,keepaspectratio]{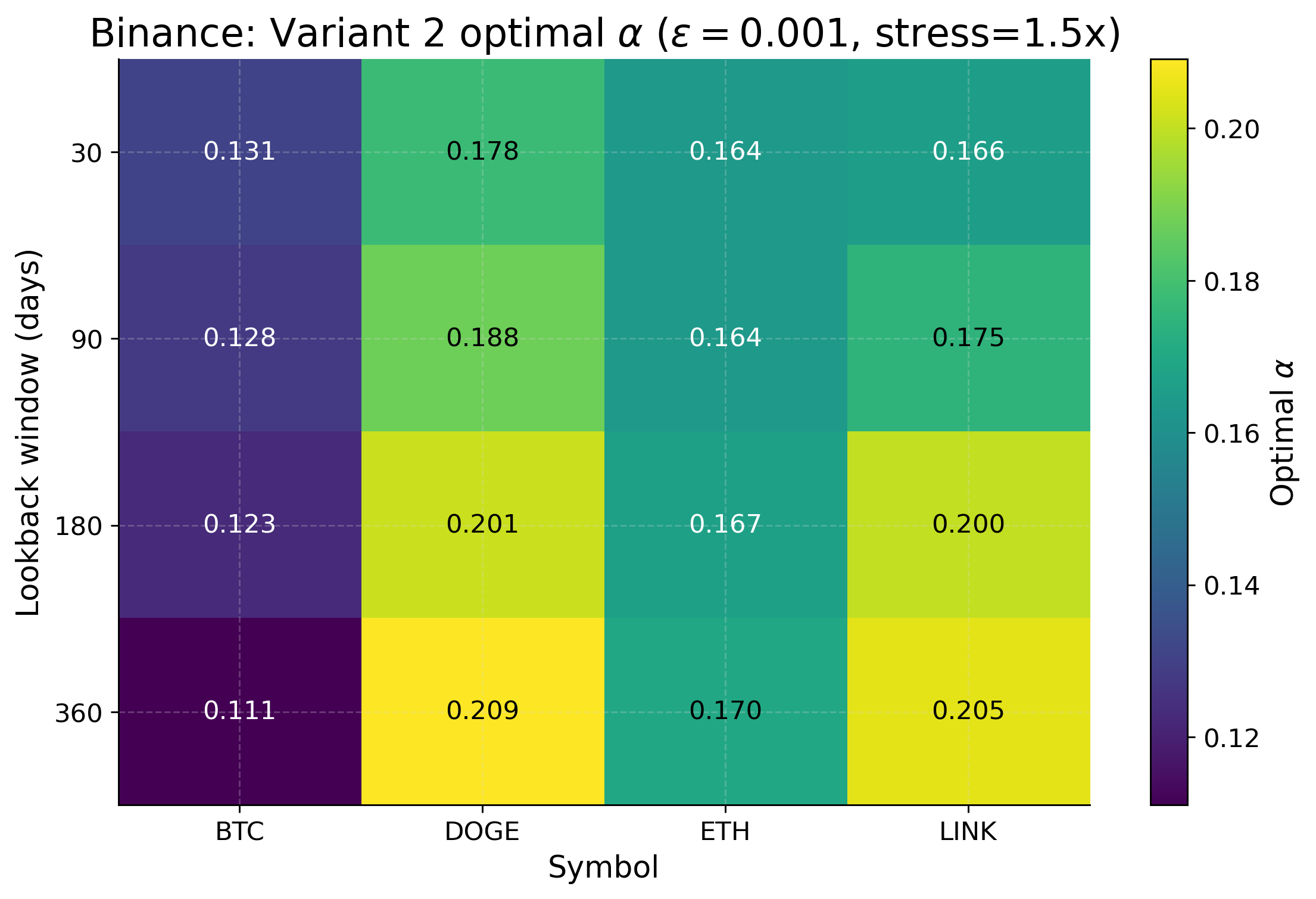}
        \caption{Binance heatmap.}
    \end{subfigure}\hfill
    \begin{subfigure}[t]{0.47\linewidth}
        \centering
        \includegraphics[width=\linewidth,height=0.13\textheight,keepaspectratio]{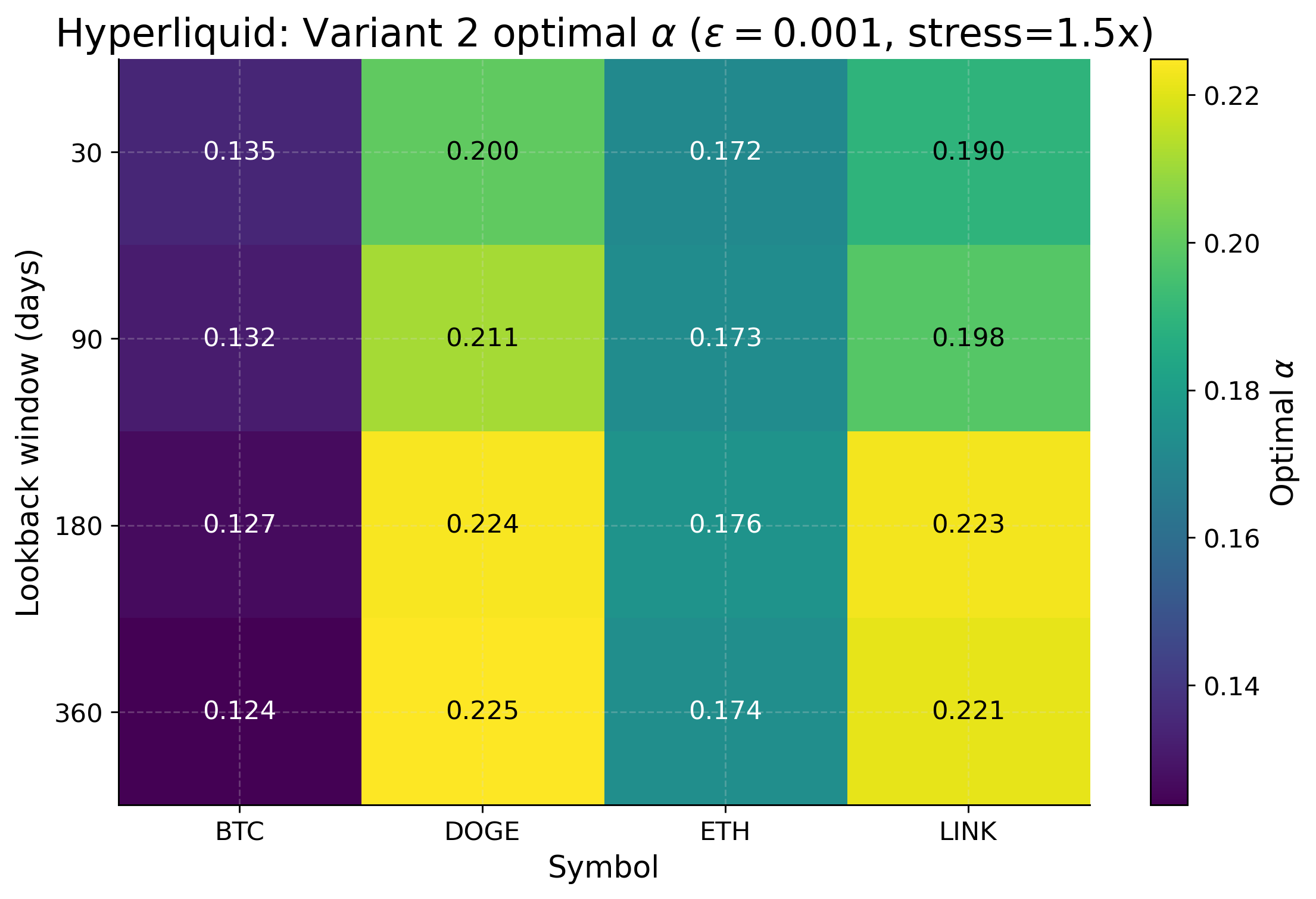}
        \caption{Hyperliquid heatmap.}
    \end{subfigure}

    \caption{Comparative simulation evidence. Top row: sensitivity curves for the benchmark 180-day, 1.5$\times$-stress slice. Bottom row: Variant~2 heatmaps for $\varepsilon=0.001$ and 1.5$\times$ stress across lookback windows.}
    \label{fig:comparative}
\end{figure}

Figure~\ref{fig:comparative} compresses the remaining comparative evidence. The upper-left panel reports Variant~2 as a function of the admissible liquidation probability; the upper-right panel reports Variant~1 as a function of $LGD$; the lower row reports venue specific heat maps for the risk-constrained benchmark family. Together, these graphs support five concise conclusions.

\paragraph*{1) The risk-constrained benchmark is the operational baseline.}
Across exchanges, assets, and calibration windows, Variant~2 is more stable than the economic optimum. It maps the solvency budget directly into a collateral rule and, therefore, provides the cleanest benchmark target for deployment.

\paragraph*{2) Volatility stress raises required collateral monotonically.}
For every asset and in both venues, a higher stressed volatility shifts the admissible control region upward. This monotonicity is one of the most robust comparative results in the paper and is exactly what one would expect from a liquidation-budget formulation.

\paragraph*{3) Long-tail assets require materially more collateral.}
BTC consistently carries the lowest benchmark collateral share, ETH lies in the middle, and LINK/DOGE require materially more conservative allocations. This is the central cross-sectional trade-off behind permissionless basis trading: the broader decentralized opportunity set is concentrated in higher beta assets that demand stricter collateralization.

\paragraph*{4) Decentralized derivative venues are modestly more conservative.}
Across the benchmark slice, the refreshed comparative calibration produces a higher optimal collateral on Hyperliquid for every asset. The difference is small for BTC and large for LINK and DOGE, which is exactly the pattern one would expect from a venue architecture that combines higher leverage dispersion, long-tail listings, and transparent margin rules. Consistent with recent evidence of \cite{ZhivkovTwoTieredFunding} that Binance leads Hyperliquid in funding rate price discovery, we treat this comparison as a reduced form venue comparison rather than as a claim that Hyperliquid is the dominant price discovery venue.

\paragraph*{5) The rolling horizon funding estimator is empirically well behaved.}
Because funding rates cluster around venue baselines while preserving a one day carry component, the rolling horizon mean is a natural estimator for the static problem. It aligns the estimator with the decision horizon and avoids the interpretational mismatch of an event level mean.

\section{Dynamic Alpha Control}

The static layer delivers the correct benchmark target, but does not solve the operational control problem. Once the position is opened, the collateral share is no longer constant: marked-to-market P\&L on the short perpetual leg and cumulative funding transfers both change the amount of capital effectively tied to derivative margin. The relevant dynamic problem is therefore to keep the realized collateral share inside an admissible band and to reset it toward a benchmark target whenever the process exits that region.

\subsection{Collateral Dynamics and Band-Control Formulation}

Recall that the marked-to-market futures account is given by \eqref{eq:futures-value}. The instantaneous collateral share may, therefore, be written as
\begin{equation}
    \alpha_t := \frac{V_t^F}{V_t},
    \qquad
    V_t = Qp_t + V_t^F,
    \label{eq:alpha-def-dyn}
\end{equation}
with $Q$ fixed between interventions. Substituting \eqref{eq:futures-value} into \eqref{eq:alpha-def-dyn} yields the full state equation
\begin{equation}
\alpha_t=
\frac{M_0 + H(f_t-f_0) - H\int_0^t \big(\iota p_r + \kappa(f_r-p_r)\big)\,dr}{Qp_t + M_0 + H(f_t-f_0) - H\int_0^t \big(\iota p_r + \kappa(f_r-p_r)\big)\,dr}.
\label{eq:alpha-full-dyn}
\end{equation}
This expression makes the mechanism explicit: $\alpha_t$ evolves endogenously because both the short-perpetual P\&L and the realized funding leg change the marked value of capital posted as margin.

\begin{figure}[t]
    \centering
    \includegraphics[width=\columnwidth]{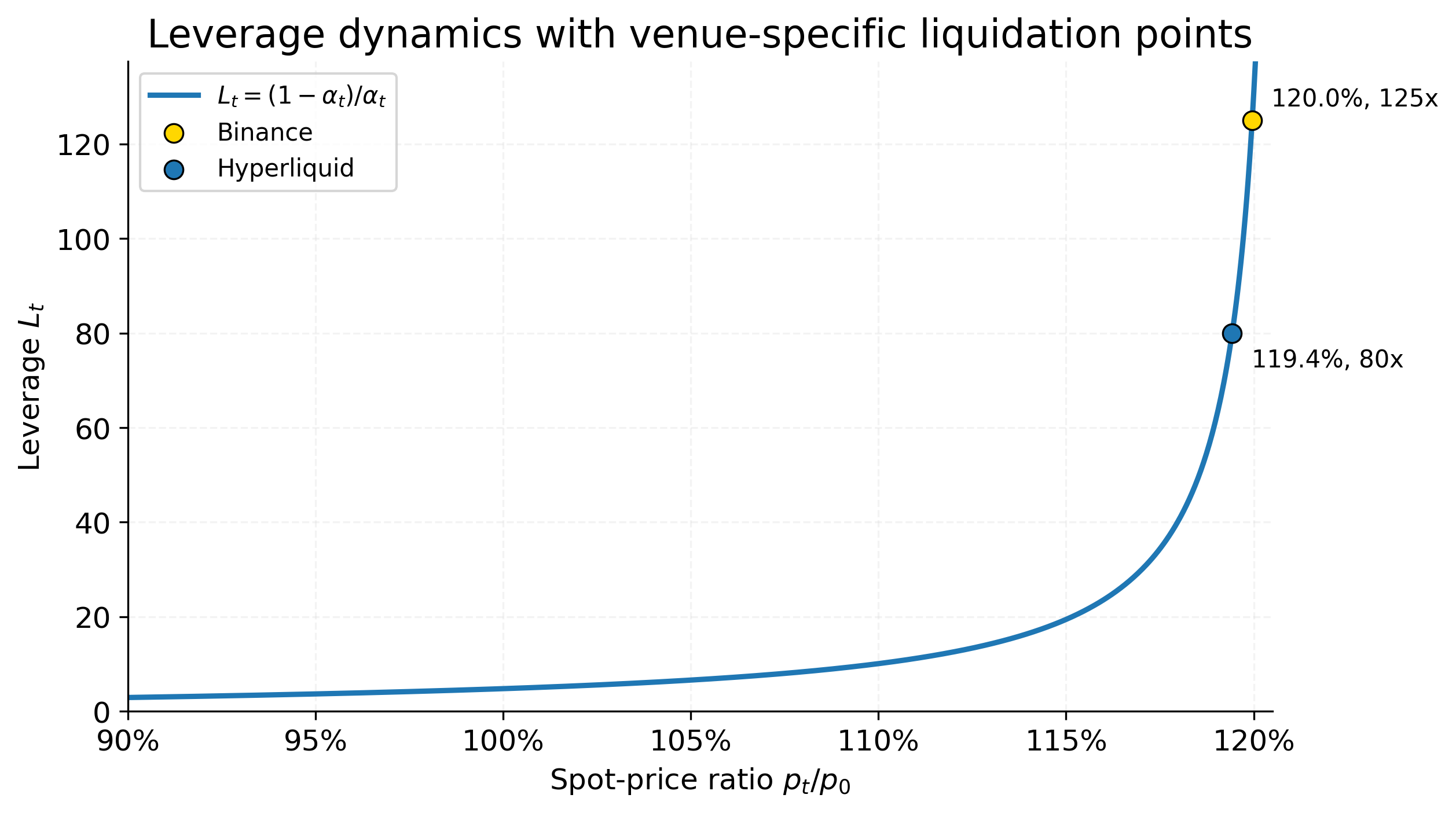}
    \caption{Reduced form hedge leg leverage dynamics implied by \eqref{eq:leverage-reduced-dyn}, shown for a fixed initial collateral share $\alpha_0$. As the spot price rises relative to its initial level, the effective collateral share declines and leverage increases nonlinearly. The colored points mark venue specific liquidation points, determined by the maintenance margin thresholds through $\alpha_{\mathrm{liq}}=\theta_F/(1+\theta_F)$. The more conservative maintenance rule liquidates earlier along the same leverage path.}
    \label{fig:leverage-dynamics}
\end{figure}

For intervention analysis, it is useful to suppress the cumulative funding term inside the state equation and work with the mark-to-market approximation alone. Under local delta neutrality and the tether $f_t=\zeta p_t$, one obtains
\begin{equation}
    \alpha_t = 1-(1-\alpha_0)\frac{p_t}{p_0},
    \label{eq:alpha-simplified-dyn}
\end{equation}
so that, under the GBM price dynamics \eqref{eq:gbm},
\begin{equation}
    d\alpha_t = -\mu(1-\alpha_t)\,dt - \sigma(1-\alpha_t)\,dW_t.
    \label{eq:alpha-sde-dyn}
\end{equation}
It is also useful to express the same reduced form dynamics in leverage terms. Define the margin leverage by
\begin{equation}
    L_t := \frac{1-\alpha_t}{\alpha_t}.
    \label{eq:leverage-def-dyn}
\end{equation}
Under \eqref{eq:alpha-simplified-dyn}, the leverage process is
\begin{equation}
    L_t = \frac{(1-\alpha_0)\frac{p_t}{p_0}}{1-(1-\alpha_0)\frac{p_t}{p_0}},
    \label{eq:leverage-reduced-dyn}
\end{equation}
with local price sensitivity,
\begin{equation}
    \frac{dL_t}{dp_t} = \frac{1-\alpha_0}{p_0\alpha_t^2}
    = \frac{1-\alpha_0}{p_0\left(1-(1-\alpha_0)\frac{p_t}{p_0}\right)^2}.
    \label{eq:leverage-sensitivity-dyn}
\end{equation}
Hence, the collateral share is itself a stochastic state variable, while the associated hedge-leg leverage is increasing and convex in the spot-price ratio under the reduced form approximation. Figure~\ref{fig:leverage-dynamics} illustrates the implied leverage map together with the venue specific liquidation points. Two features are immediate. First, upward spot moves compress the effective margin buffer and generate a strongly nonlinear increase in leverage. Second, liquidation occurs strictly before the singular limit $\alpha_t=0$: for a given initial collateral share, the relevant operational thresholds are the venue specific points at which $\alpha_t$ reaches
\[
\alpha_{\mathrm{liq}}=\frac{\theta_F}{1+\theta_F},
\]
rather than the unattained boundary $\alpha_t=0$. The figure therefore highlights both the convexity of the leverage map and the fact that tighter maintenance rules imply earlier liquidation along the same reduced form trajectory.

We therefore formulate the dynamic layer as a band-control problem. Let
\begin{equation}
    \mathcal{A}:=[\alpha_L,\alpha_U]
\end{equation}
be the admissible region and let $\alpha^{\dagger}\in(\alpha_L,\alpha_U)$ denote the reset target. The controller keeps $\alpha_t$ inside $\mathcal{A}$ and, after leaving the band, restores the position toward $\alpha^{\dagger}$. In reduced form, the policy is as follows:
\begin{equation}
\alpha_t\in[\alpha_L,\alpha_U]\Rightarrow \text{no action},
\qquad
\alpha_t\notin[\alpha_L,\alpha_U]\Rightarrow \alpha_t\mapsto \alpha^{\dagger}.
\label{eq:band-policy-dyn}
\end{equation}
We take $\alpha^{\dagger}$ from the static problem constrained by the benchmark risk in Table~\ref{tab:main}. This choice is economically natural: the static constrained solution already converts the liquidation budget into an interpretable collateral rule, so the dynamic layer should stabilize that benchmark rather than replace it with a second, unrelated target.

\subsection{Lower Intervention Boundary from Operational Liquidation Control}

The lower intervention boundary appears by construction and is solvency driven. Let $h_{\mathrm{liq}}>0$ denote the short operational horizon over which the trading agent must remain solvent without a completed rebalance, and let $\varepsilon_{\mathrm{liq}}\in(0,1)$ denote the acceptable liquidation probability over that interval. We define
\begin{equation}
    \alpha_L := \inf\{\alpha\in(0,1):\Pi_{\mathrm{liq}}(\alpha;h_{\mathrm{liq}})\le \varepsilon_{\mathrm{liq}}\}.
    \label{eq:alpha-lower-dyn}
\end{equation}
The parameter $h_{\mathrm{liq}}$ should be interpreted as an execution latency horizon rather than as a portfolio holding period. In the implementation, it is generated by the reaction time of the trading agent, DEX routing delay, collateral-transfer latency, and related operational frictions. We defer its final calibration to the execution constraints discussion, but the logic is already clear: the lower boundary is pinned down by a short horizon liquidation budget and, therefore, remains meaningful independently of the economics of downward rebalancing. A natural first-order adaptive extension would replace the fixed lower boundary by a state-dependent threshold
\begin{equation}
    \alpha_L(t)
    :=
    \inf\left\{
    \alpha \in (0,1):
    \Pi_{\mathrm{liq}}(\alpha; h_{\mathrm{liq}}, \sigma_t)\le \varepsilon_{\mathrm{liq}}
    \right\},
    \label{eq:alpha-lower-adaptive}
\end{equation}
where $\sigma_t$ is an estimate of the short horizon local volatility. In the present paper, however, we retain a conservative fixed approximation $\alpha_L$ in order to keep the control rule interpretable and the backtest layer ex ante and frozen.

\subsection{Bootstrap Robustness of the Lower Boundary}

The diffusion benchmark is deliberately parsimonious, which is analytically useful for deriving transparent control rules but may smooth short horizon tail behavior too aggressively for crypto assets. To strengthen the solvency interpretation of the lower boundary without abandoning the benchmark model, we therefore perform a bootstrap robustness exercise on the refreshed comparative calibration. The goal is not to replace the GBM-based control architecture, but to verify that the lower intervention logic remains conservative when the short-run return law is taken directly from historical data rather than from Gaussian shocks.

For each venue and asset, we take the refreshed 180-day benchmark window and collect one hour historical log returns. We then resample these short horizon returns with replacement to form synthetic three-hour paths, matching the operational liquidation horizon used in the dynamic layer. To preserve the conservative calibration philosophy used elsewhere in the paper, resampled returns are additionally stressed by a $1.5\times$ volatility multiplier, and we set
\[
h_{\mathrm{liq}}=3\text{ hours},
\qquad
\varepsilon_{\mathrm{liq}}=10^{-4}.
\]
The rationale for the choice of $h_{\mathrm{liq}}$ is discussed later in Section IX.A. The liquidation budget $\varepsilon_{\mathrm{liq}}$ is used here as a deliberately strict conservative bootstrap design parameter for the lower-bound robustness exercise.
For each venue--asset pair, the bootstrap lower boundary is defined as
\begin{equation}
    \alpha_L^{\mathrm{boot}}
    :=
    \inf\left\{\alpha\in(0,1):
    \widehat{\Pi}_{\mathrm{liq}}^{\mathrm{boot}}(\alpha;h_{\mathrm{liq}})
    \le \varepsilon_{\mathrm{liq}}\right\},
    \label{eq:alpha-lower-bootstrap}
\end{equation}
where $\widehat{\Pi}_{\mathrm{liq}}^{\mathrm{boot}}$ is the empirical liquidation probability estimated from the resampled short horizon paths.

\begin{figure}[t]
    \centering
    \includegraphics[width=\columnwidth]{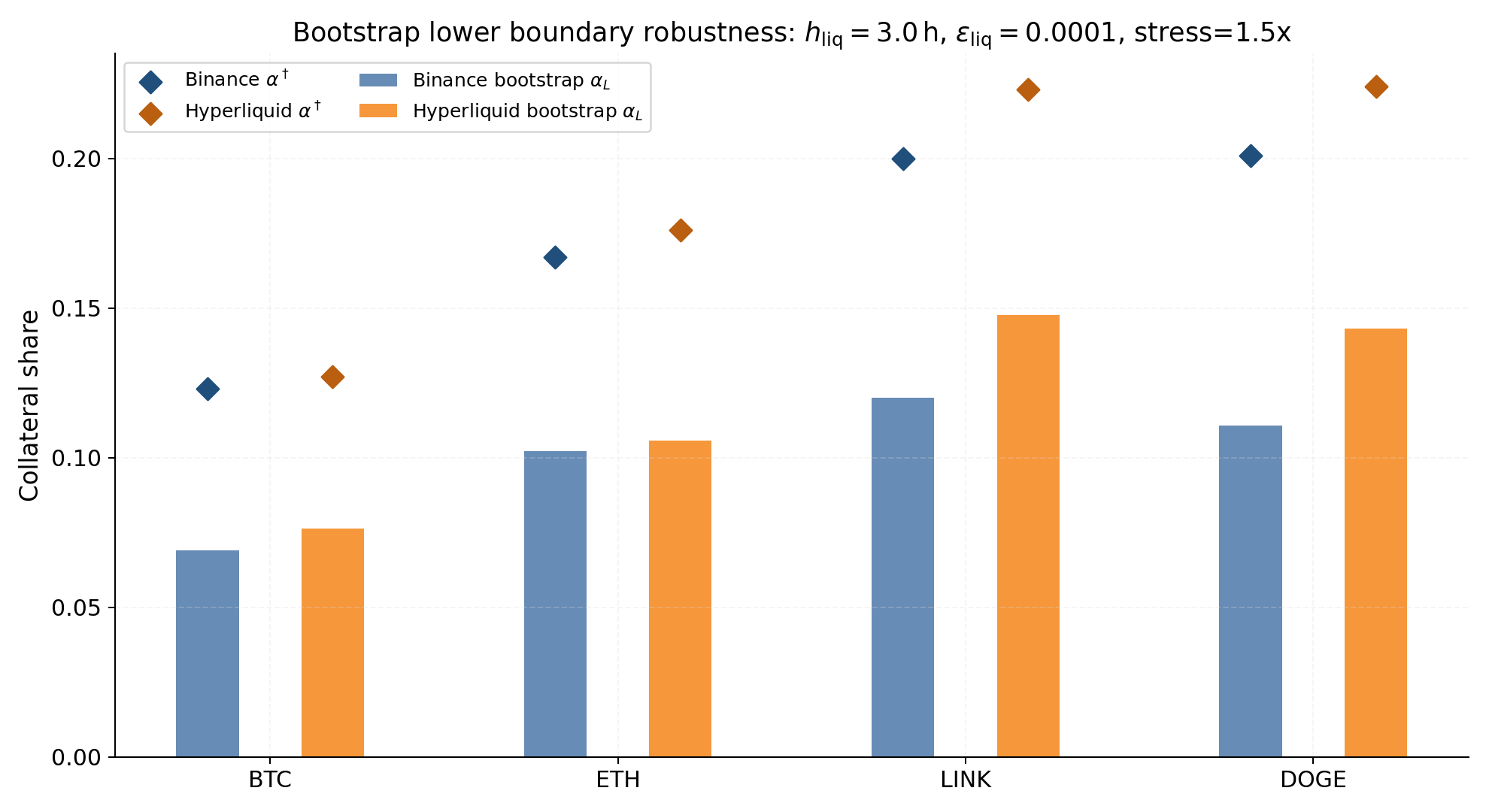}
    \caption{Bootstrap robustness of the dynamic lower boundary on the refreshed comparative calibration. Bars report the bootstrap-based lower boundary $\alpha_L^{\mathrm{boot}}$, while diamonds report the benchmark target $\alpha^{\dagger}$. Parameters: one hour return bootstrap, 180-day lookback, $h_{\mathrm{liq}}=3$ hours, $\varepsilon_{\mathrm{liq}}=10^{-4}$, and $1.5\times$ volatility stress.}
    \label{fig:alpha-lower-bootstrap}
\end{figure}

\begin{table}[t]
\centering
\caption{Bootstrap lower-bound robustness on the refreshed comparative calibration. The bootstrap uses one hour returns from the refreshed 180-day calibration window, a three-hour liquidation horizon, $\varepsilon_{\mathrm{liq}}=10^{-4}$, and $1.5\times$ volatility stress.}
\label{tab:fresh-bootstrap-lower}
\small
\begin{tabular}{llccc}
\toprule
Venue & Asset & $\alpha^{\dagger}$ & $\alpha_L^{\mathrm{boot}}$ & $\alpha^{\dagger}-\alpha_L^{\mathrm{boot}}$ \\
\midrule
Binance & BTC  & 0.123 & 0.069 & 0.054 \\
Binance & ETH  & 0.167 & 0.102 & 0.065 \\
Binance & LINK & 0.200 & 0.120 & 0.080 \\
Binance & DOGE & 0.201 & 0.111 & 0.090 \\
Hyperliquid & BTC  & 0.127 & 0.076 & 0.051 \\
Hyperliquid & ETH  & 0.176 & 0.106 & 0.070 \\
Hyperliquid & LINK & 0.223 & 0.148 & 0.075 \\
Hyperliquid & DOGE & 0.224 & 0.143 & 0.081 \\
\bottomrule
\end{tabular}
\end{table}

Figure~\ref{fig:alpha-lower-bootstrap} yields three conclusions, and Table~\ref{tab:fresh-bootstrap-lower}. First, the qualitative ranking remains unchanged: BTC remains the least collateral intensive asset, ETH lies in the middle, and LINK/DOGE remain the most demanding assets under both venues. Second, the decentralized benchmark remains more conservative in the bootstrap exercise as well, with Hyperliquid producing a higher empirical lower boundary than Binance for every studied asset. Third, and most importantly, the benchmark target $\alpha^{\dagger}$ remains materially above the bootstrap lower boundary across all assets and at both venues. The dynamic target should therefore be interpreted as an operational collateral target rather than as a knife-edge solvency floor. In this sense, the bootstrap exercise strengthens the paper's main control result: even after allowing the short horizon return law to inherit empirical tail behavior from the data, the lower intervention logic continues to imply a meaningful solvency buffer rather than a minimally feasible margin allocation.

\subsection{Upper Intervention Boundary from Carry Loss and Rebalancing Costs}

The upper intervention boundary is economic rather than solvency-driven. Let $\kappa_h$ denote the cumulative funding return over a decision horizon $h$, and define
\[
\widetilde{\kappa}_h := \mathbb{E}[\kappa_h].
\]
For a fixed collateral share $\alpha$, define the expected carry over horizon $h$ by
\begin{equation}
    \mathcal{C}(\alpha;h)
    :=
    \mathbb{E}\!\left[(1-\alpha)D\,\kappa_h\right]
    =
    (1-\alpha)D\,\widetilde{\kappa}_h.
    \label{eq:carry-alpha-dyn}
\end{equation}
Rebalancing from a state $\alpha>\alpha^{\dagger}$ back to the target is worthwhile whenever the expected carry recovered by freeing the collateral exceeds the effective rebalancing cost. Thus,
\begin{equation}
    \mathcal{C}(\alpha^{\dagger};h)-\mathcal{C}(\alpha;h)\ge K_{\mathrm{reb}}.
    \label{eq:upper-indifference-1}
\end{equation}
Since
\begin{align}
    \mathcal{C}(\alpha^{\dagger};h)-\mathcal{C}(\alpha;h)
    &= \big[(1-\alpha^{\dagger})-(1-\alpha)\big]D\widetilde{\kappa}_h \notag\\
    &= (\alpha-\alpha^{\dagger})D\widetilde{\kappa}_h,
    \label{eq:upper-indifference-2}
\end{align}
the upper boundary solves
\begin{equation}
    (\alpha_U-\alpha^{\dagger})D\widetilde{\kappa}_h = K_{\mathrm{reb}},
    \label{eq:alpha-upper-dyn}
\end{equation}
and therefore
\begin{equation}
    \Delta_U := \alpha_U-\alpha^{\dagger} = \frac{K_{\mathrm{reb}}}{D\widetilde{\kappa}_h}.
    \label{eq:delta-upper-dyn}
\end{equation}
The rebalancing cost term is broadly interpreted as
\begin{equation}
    K_{\mathrm{reb}} = K_{\mathrm{fee}} + K_{\mathrm{impact}} + K_{\mathrm{gas}} + K_{\mathrm{exec}} - B_{\mathrm{basis}},
    \label{eq:kreb-structure-dyn}
\end{equation}
where $K_{\mathrm{fee}}$ is the spot swap fee, $K_{\mathrm{impact}}$ is the impact and routed slippage of the price, $K_{\mathrm{gas}}$ is the blockchain execution cost, $K_{\mathrm{exec}}$ collects the latency related implementation frictions, and $B_{\mathrm{basis}}$ denotes any favorable basis contribution realized at the moment of intervention. A useful validation case follows immediately: if $K_{\mathrm{reb}}\le 0$, then downward rebalancing is always optimal, since the intervention itself is economically favorable.

\begin{figure}[t]
    \centering
    \includegraphics[width=\columnwidth]{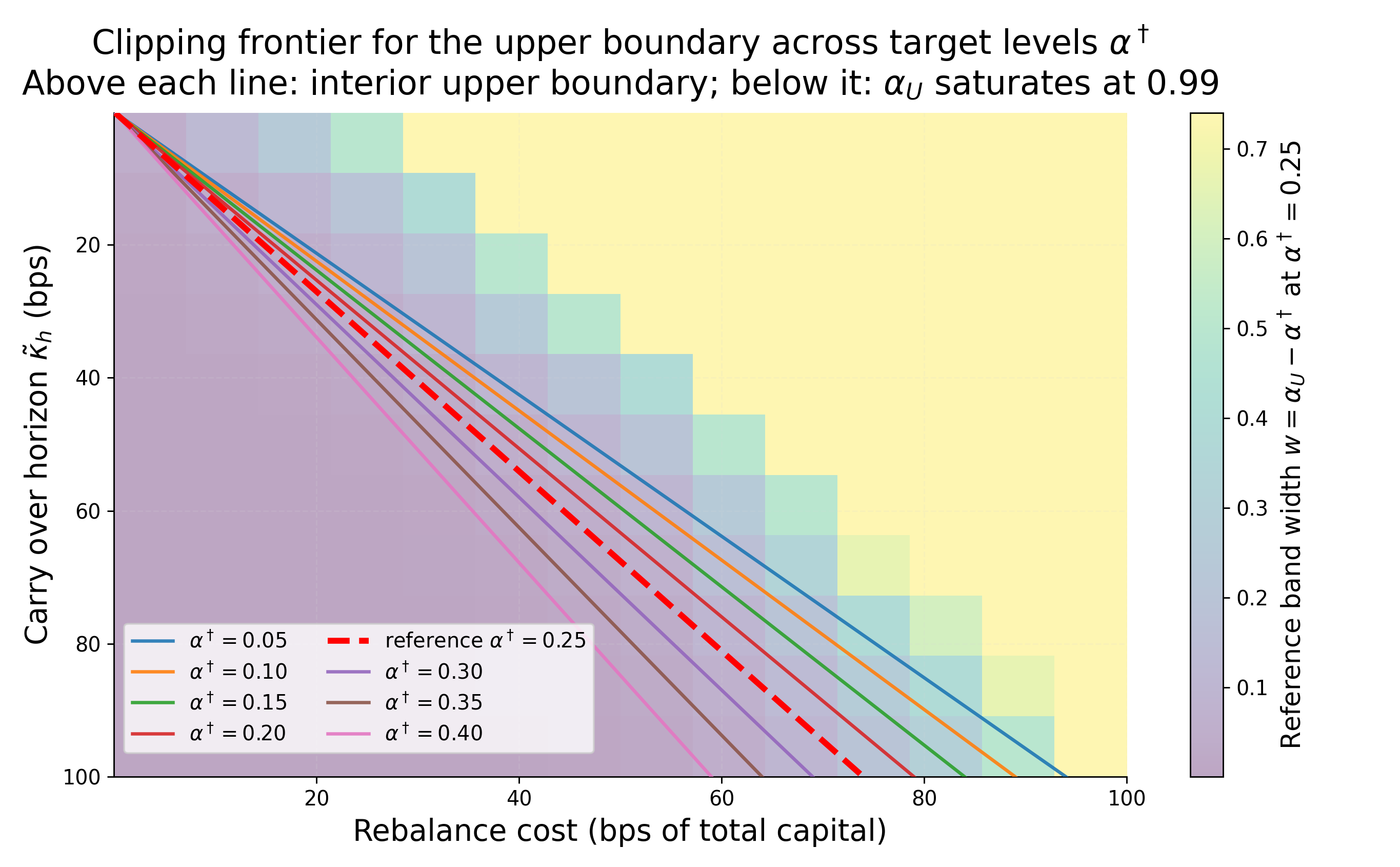}
    \caption{Upper-band width $\Delta_U=\alpha_U-\alpha^{\dagger}$ and clipping frontiers across target collateral levels. Higher benchmark targets leave less room for an interior upper boundary before saturation at $\alpha_U=0.99$.}
    \label{fig:alpha-band}
\end{figure}

Figure~\ref{fig:alpha-band} reports the implied upper-band width $\Delta_U$ together with the clipping frontier between the benchmark targets. Two conclusions are central. First, the clipping lines shift inward as $\alpha^{\dagger}$ increases: a more conservative target leaves less room before the upper boundary saturates at one, so the admissible interior region for a meaningful upper trigger shrinks. Second, under realistic DeFi execution cost assumptions in the 10--30 bps range, the economically meaningful width $\Delta_U$ is typically coarse, roughly in the 0.2--0.5 range wherever an interior solution still exists. Hence a purely funding-driven upper trigger is intrinsically blunt even before one models execution microstructure in full detail.

The remaining difficulty is in selecting the horizon. Equation~\eqref{eq:delta-upper-dyn} requires an estimate of the expected funding over a decision horizon $h$, but that horizon is not known a priori. If chosen too long, the expected carry is overstated and downward rebalancing is triggered too aggressively. We therefore examine the persistence of candidate upper triggers directly through Monte Carlo evidence.

\subsection{Monte Carlo Evidence on Rebalancing Lifetime}

\begin{figure}[t]
    \centering
    \includegraphics[width=\columnwidth]{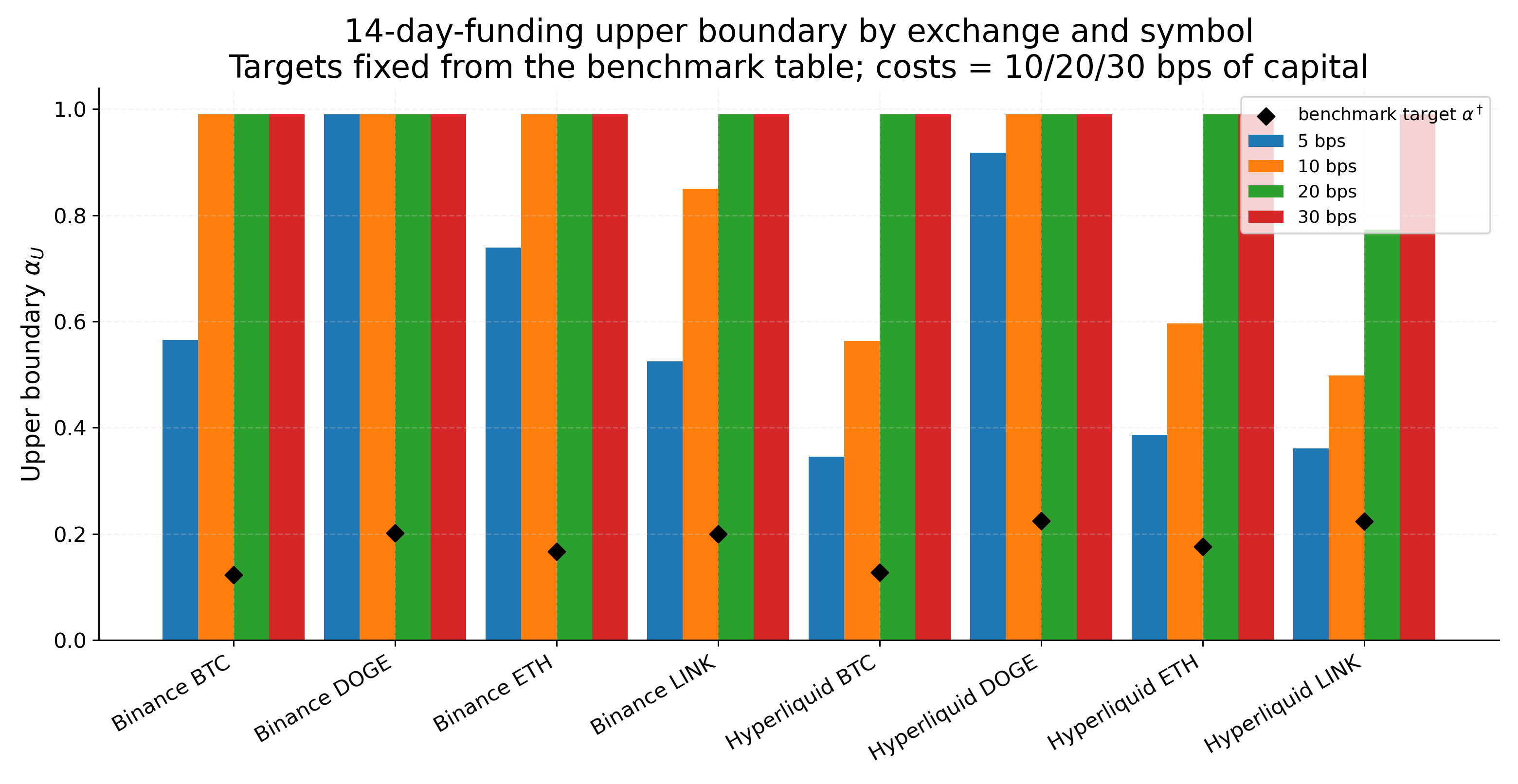}
    \caption{Monte Carlo upper-bound evidence across exchanges and symbols. Black diamonds mark the benchmark targets from the static constrained problem; colored bars report the implied upper boundary under alternative rebalance-cost scenarios.}
    \label{fig:alpha-upper-mc}
\end{figure}

We simulate GBM price paths under the benchmark volatility calibration and evaluate the time required to reach candidate upper boundaries implied by a 14-day expected-funding estimate. The reset targets are fixed on the static benchmark slice in Table~\ref{tab:main}, and the rebalancing cost scenarios are 5, 10, 20 and 30 bps of capital. Figure~\ref{fig:alpha-upper-mc} reports the resulting implied upper boundaries, while Appendix~C collects the corresponding numerical summary.

The Monte Carlo evidence yields several strong conclusions. Meaningful interior values of $\alpha_U$ appear mainly in the low cost regime, roughly around 5 bps. They are more plausible in richer funding environments, which partially confirms the idea that permissionless basis trading is most attractive on markets with high funding carry. The effect is most visible on Hyperliquid, and especially for LINK, with ETH and BTC staying behind. Similarly, upper control is venue specific, asset specific, and not universal. For realistic DeFi costs in the 10--30 bps range, many cases already saturate at $\alpha_U\approx 0.99$, which means that a funding-only upper trigger ceases to be operationally relevant.

\subsection{Practical Rule-Based Upper Control}

This motivates a practical rule-based interpretation of the upper boundary. In implementation, the controller fixes an execution cost budget and applies \eqref{eq:alpha-upper-dyn} as a policy rule rather than as a continuously re-estimated state-dependent control law. In operational form, downward rebalancing is triggered only when both conditions hold
\begin{equation}
\left\{
\begin{aligned}
\alpha_t &> \alpha_U,\\
(\alpha_t-\alpha^{\dagger})D\widetilde{\kappa}_h &\ge K_{\mathrm{reb}}^{\mathrm{fix}}.
\end{aligned}
\right.
\label{eq:upper-trigger-system}
\end{equation}
Here $K_{\mathrm{reb}}^{\mathrm{fix}}$ is the fixed operational execution cost budget used by the controller. This is preferable at the present stage because the basis component in \eqref{eq:kreb-structure-dyn} can materially alter the economics of the intervention carried out from one episode to the next. If the basis spread moves in the operator's favor, then the net rebalancing cost can become non-positive, in which case rebalancing is always economically justified. The lower boundary remains naturally tied to short horizon solvency and, therefore, admits a cleaner model based interpretation than the upper boundary.

Our practical continuous control specification is therefore the following. The target $\alpha^{\dagger}$ is fixed at the static optimum constrained by the benchmark. The lower boundary $\alpha_L$ is determined by the short horizon liquidation criterion \eqref{eq:alpha-lower-dyn}, with $h_{\mathrm{liq}}$ later calibrated from execution latency and trading agent delays. The upper boundary $\alpha_U$ is generated from the carry-loss compared to the cost \eqref{eq:alpha-upper-dyn}, but in implementation it should be treated as a coarse operational cap whose relevance depends on venue, asset, funding regime and execution costs.

Continuous alpha control is therefore asymmetrical by construction. The lower trigger is structural and solvency driven. The upper trigger is economic, regime dependent, and may disappear entirely under realistic execution costs. The resulting blueprint is operationally coherent: a benchmark target inherited from the static constrained problem, a liquidation-budget lower boundary, and a rule based upper intervention boundary filtered through explicit execution costs.

\subsection{Operational Control Algorithm}

The preceding analysis yields an asymmetric operational rule. The lower boundary is interpreted as a solvency-first intervention threshold, whereas the upper boundary is interpreted as an economic trigger that is applied only after solvency is preserved. In particular, lower-bound breaches are handled through an immediate buy-side rebalancing logic, while upper-bound breaches require an additional carry-versus-cost check before any downward rebalancing is executed. The minimum executable notional $q_{\min}$ and the execution buffer $b$ will be discussed later in the execution Section~X.

Implementation also requires a distinction between the rebalancing size that would restore the portfolio toward the benchmark target and the size that is actually executable under the prevailing cost budget. Accordingly, the operational rule first computes a target rebalancing size and then truncates it by the contemporaneously feasible executable size. This implies that interventions reset the state only towards $\alpha^{\dagger}$ and do not need to restore the target exactly when execution costs are binding. These steps are summarized in Table~\ref{tab:alpha-control-algorithm}.

\begin{table}[t]
\centering
\caption{Operational alpha-control policy used to interpret the dynamic layer.}
\label{tab:alpha-control-algorithm}
\footnotesize
\setlength{\tabcolsep}{4pt}
\begin{tabular}{p{0.97\columnwidth}}
\toprule
\textbf{Inputs:} benchmark target $\alpha^{\dagger}$; lower boundary $\alpha_L$ from \eqref{eq:alpha-lower-dyn}; upper boundary $\alpha_U$ from \eqref{eq:alpha-upper-dyn}; expected funding estimate $\widetilde{\kappa}_h$ over decision horizon $h$; fixed execution cost budget $K_{\mathrm{reb}}^{\mathrm{fix}}$; minimum executable notional $q_{\min}$; side-specific executable cost function $C_s(q,t)$ for $s\in\{\mathrm{buy},\mathrm{sell}\}$. \\
\midrule
\textbf{1. Update state.} Observe spot and perpetual prices and update $V_t^F$, $V_t$, and $\alpha_t = V_t^F/V_t$. \\
\addlinespace[0.25em]
\textbf{2. Classify the region.} 
\[
\alpha_t < \alpha_L \Rightarrow s=\mathrm{buy},\qquad
\alpha_t > \alpha_U \Rightarrow s=\mathrm{sell}.
\]
If $\alpha_t\in[\alpha_L,\alpha_U]$, do not rebalance. \\
\addlinespace[0.25em]
\textbf{3. Compute target and executable size.} For the active side $s$, compute the target rebalancing size $q_s^\star(t)$ implied by restoring the position toward $\alpha^{\dagger}$. Then compute
\[
Q_s^{\max}(t):=\sup\{q\ge 0:\, C_s(q,t)\le K_{\mathrm{reb}}^{\mathrm{fix}}\},
\]
\[
\widehat q_s(t):=\min\{q_s^\star(t),\,Q_s^{\max}(t)\}.
\]
\\
\addlinespace[0.25em]
\textbf{4. Execute lower-bound intervention.} If $\alpha_t < \alpha_L$ and $\widehat q_{\mathrm{buy}}(t)\ge q_{\min}$, execute the buy-side rebalancing of size $\widehat q_{\mathrm{buy}}(t)$. If $\alpha_t < \alpha_L$ but $\widehat q_{\mathrm{buy}}(t)< q_{\min}$, escalate to the emergency execution path. \\
\addlinespace[0.25em]
\textbf{5. Execute upper-bound intervention.} If $\alpha_t > \alpha_U$, first check
\[
(\alpha_t-\alpha^{\dagger})D\widetilde{\kappa}_h \ge K_{\mathrm{reb}}^{\mathrm{fix}}.
\]
Execute the sell-side rebalancing only if this condition holds and $\widehat q_{\mathrm{sell}}(t)\ge q_{\min}$. Otherwise, keep the position unchanged. \\
\bottomrule
\end{tabular}
\end{table}

\section{Backtest Calibration and Historical Validation}

\subsection{Historical Calibration for the Backtest Layer}

Before live deployment, the control rule was calibrated in an older historical window and then frozen for out-of-time validation. In this layer, we use Binance price and funding data over 2023-01-01 to 2024-12-31 as the historical market proxy, while retaining Hyperliquid's target margin rule $\theta_F=1/(2L_{\max})$ and Hyperliquid maximum leverage for each asset in order to mirror the original ex ante design problem. Table~\ref{tab:backtest-benchmark-slice} reports the resulting benchmark constrained targets for the backtest layer, and Appendix~D reports the extended static grids and Monte Carlo band inputs.

\begin{table}[t]
\centering
\caption{Historical benchmark slice used to initialize the backtest layer. The data window is Binance 2023-01-01 to 2024-12-31; the margin rule is the current Hyperliquid rule $\theta_F=1/(2L_{\max})$. Variant~2, $\varepsilon=0.001$, $1.5\times$ volatility stress, 180d lookback.}
\label{tab:backtest-benchmark-slice}
\small
\setlength{\tabcolsep}{4pt}
\begin{tabular}{lccccc}
\toprule
Asset & $\alpha^{\dagger}$ & Lev. & $\widetilde{\kappa}_{1d}$ (bps) & $\theta_F$ & $L_{\max}$ \\
\midrule
BTC  & 0.173 & 4.78x & 2.19 & 0.0125 & 40 \\
ETH  & 0.183 & 4.46x & 2.56 & 0.0200 & 25 \\
LINK & 0.283 & 2.53x & 2.63 & 0.0500 & 10 \\
DOGE & 0.299 & 2.34x & 2.60 & 0.0500 & 10 \\
\bottomrule
\end{tabular}
\end{table}

For the lower boundary in the backtest layer, we calibrate the liquidation horizon conservatively. The strategy is intended for Arbitrum because this chain gives direct access to the on-chain setup while also providing a direct bridge path into Hyperliquid's settlement environment \cite{HyperliquidBridge,HyperliquidBridge2}. In normal conditions, the operational execution latency is short, but publicly documented sequencer interruptions on Arbitrum have lasted materially longer than the baseline transaction-processing horizon, including a roughly 45-minute outage in September 2021 and a longer disruption in December 2023 \cite{ArbitrumOutage2021,ArbitrumOutage2023}. Since the stressed interruption horizon dominates the normal operational latency scale, the historical backtest layer uses
\begin{equation}
    h_{\mathrm{liq}}=3\text{ hours},
    \qquad
    \varepsilon_{\mathrm{liq}}=10^{-4},
\end{equation}
which is stricter than the static baseline used for the benchmark target. Historical backtests were run with the Fractal research framework \cite{FractalDeFiStrategy}, using two assumptions to simplify the backtesting processes that focus on the control framework  1) the fixed execution costs schedule in Table~\ref{tab:backtest-exec-costs} and 2) instant rebalancing liquidity with that fixed cost. The strategy was evaluated to validate the robustness of the control rule rather than to optimize a single point estimate, and we translate the collateral share into hedge-leg leverage according to $\mathrm{L}(\alpha)=(1-\alpha)/\alpha$.

\begin{table}[b]
    \centering
    \caption{Fixed execution costs used in the historical backtest engine.}
    \label{tab:backtest-exec-costs}
    \small
    \begin{tabular}{lc}
        \toprule
        Asset & Execution cost (bps) \\
        \midrule
        BTC   & 20 \\
        ETH   & 20 \\
        DOGE  & 30 \\
        LINK  & 30
    \end{tabular}
\end{table}

\subsection{Historical Validation under Alternative Funding Environments}

Building on the baseline backtest of Section~IX.A, we now conduct a controlled funding-environment experiment: holding the calibrated control rule, the Hyperliquid margin architecture, and the Binance price data all fixed, we replace the Binance funding stream with the contemporaneous Hyperliquid funding stream. The historical validation layer thereby serves three distinct purposes. First, it checks whether the present control based formulation is economically consistent with the previous Logarithm Labs  research \cite{BasisOSWhitepaper} rather than a qualitatively different strategy design. Second, it validates the historically calibrated frozen policy out of time in 2024--2025. Third, by holding the calibrated control rule and target margin architecture fixed while varying only the funding stream supplied to the hedge leg, it isolates the economic role of the funding environment itself.

The first claim is the result of the consistency of the methods. The older study was optimized directly over leverage bands, while the present paper calibrates a collateral share control and then maps it to leverage. However, the historical implied target leverage in Table~\ref{tab:backtest-benchmark-slice} remains close to the high-performing regions reported in the earlier grid search: BTC and ETH continue to support materially higher target leverage (leverage bound is 4-6) than long-tail assets, while DOGE remains economically viable only in substantially more conservative leverage regimes (leverage bound is 2-4) \cite{BasisOSWhitepaper}. The new control formulation should therefore be interpreted as a sharper and more execution-aware restatement of the earlier empirical intuition rather than as a contradictory redesign.

The second claim is an empirical validation of the frozen calibrated policy. Table~\ref{tab:backtest-funding-summary} shows that the historically calibrated rule remains operationally coherent out of time across all four assets: drawdowns remain small, realized hedge-leg leverage stays well below venue caps, and rebalancing counts remain modest over the full validation year. Table~\ref{tab:backtest-rebalance-diagnostics} complements this with direct diagnostics of policy activities. The rule does not overtrade: BTC requires only five rebalances under both funding environments, ETH and LINK require six to seven, and DOGE --- the highest-carry and highest-beta case --- remains below ten interventions even though it is the most active of the four assets. This pattern is exactly what the control logic predicts: DOGE offers the strongest economic carry but requires the most active management.

\begin{figure}[t]
    \centering
    \includegraphics[width=\linewidth]{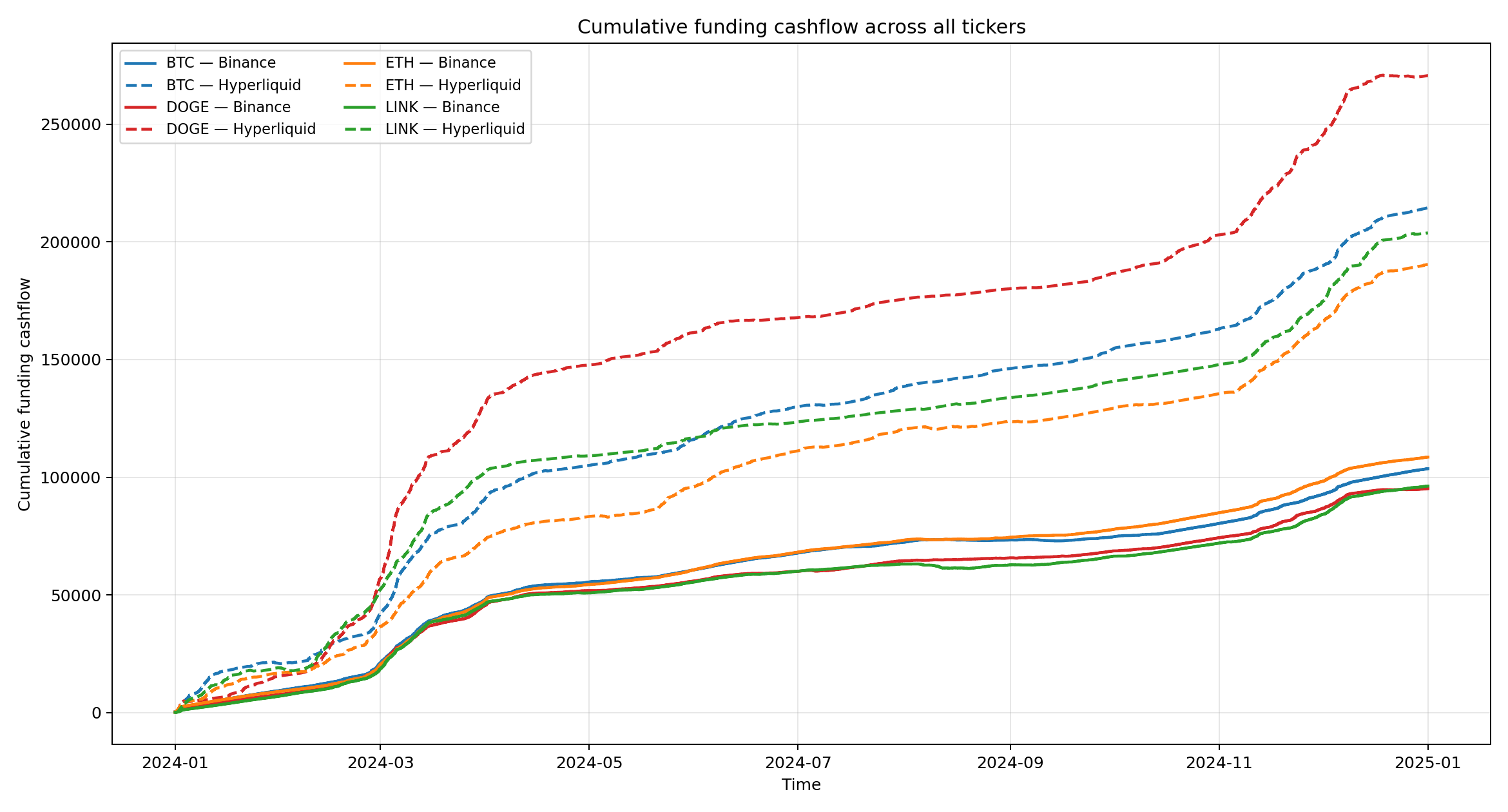}
    \caption{Historical validation under alternative funding environments. The calibrated control rule and Hyperliquid margin architecture are held fixed, while only the funding input is changed. The realized cross-sectional gap is therefore attributable to the carry environment rather than to a change in the control rule itself.}
    \label{fig:all-tickers-funding-history}
\end{figure}

The third claim concerns funding itself. Table~\ref{tab:backtest-funding-delta}, Figure~\ref{fig:all-tickers-funding-history} and Figure~\ref{fig:all-tickers-funding-cashflow} show that once the control rule is held fixed, the funding environment materially changes the realized strategy economics across every asset studied. Hyperliquid funding produces a positive increase in the APY in all four cases, ranging from roughly $8.2\%$ for ETH to $17.6\%$ for DOGE. Just as importantly, APY and funding APY remain very close in both environments. The realized strategy therefore behaves predominantly as a basis-carry policy rather than as a directional trading rule. DOGE is the clearest high-carry case: it produces the largest funding uplift and the highest total APY under Hyperliquid funding, but it also remains the most active control regime, which is consistent with its higher volatility and tighter solvency-management requirements.

\begin{figure}[t]
    \centering
    \includegraphics[width=\linewidth]{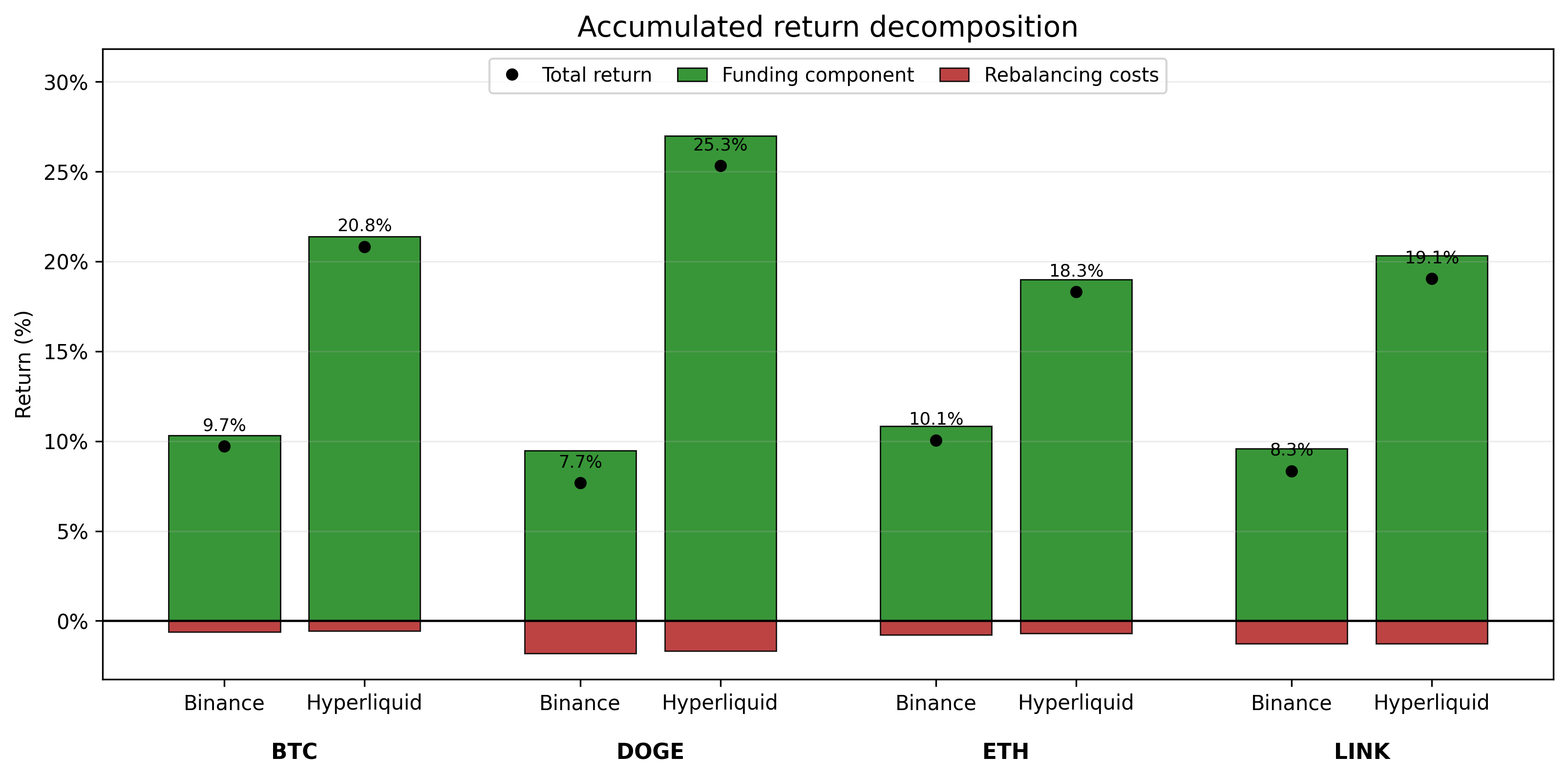}
    \caption{Historical validation under alternative funding environments. Bars report accumulated return decomposition during the backtest period, with funding conditions summarized by net funding APY.}
    \label{fig:return-decomposition}
\end{figure}

Then, according to the assumptions of the present backtest, the Figure~\ref{fig:return-decomposition} funding is the dominant positive return component, while rebalancing costs remain comparatively small and negative. Hyperliquid consistently delivers higher total returns than Binance, suggesting that the cross-venue performance gap is driven primarily by the stronger funding environment rather than by differences in cost drag. Whether the same pattern survives under live execution is an empirical question addressed separately in the execution-validation layer.

\begin{table}[t]
\centering
\caption{Ticker-level change when moving from Binance funding to Hyperliquid funding under the same frozen control rule.}
\label{tab:backtest-funding-delta}
\small
\setlength{\tabcolsep}{5pt}
\begin{tabular}{lcccc}
\toprule
Ticker & $\Delta$ APY (\%) & $\Delta$ Funding APY (\%) & $\Delta$ Avg leverage \\
\midrule
BTC  & 11.06 & 11.05 & -0.52 \\
DOGE & 17.58 & 17.50 & -0.19 \\
ETH  & 8.22  & 8.16  & -0.37 \\
LINK & 10.67 & 10.72 & -0.23 \\
\bottomrule
\end{tabular}
\end{table}

\begin{table*}[t]
\centering
\caption{Historical backtest summary under alternative funding environments with fixed historically calibrated control parameters and fixed Hyperliquid margin architecture.}
\label{tab:backtest-funding-summary}
\small
\setlength{\tabcolsep}{5pt}
\begin{tabular}{llcccccc}
\toprule
Ticker & Funding environment & Acc. return (\%) & APY (\%) & Funding APY (\%) & Max DD (\%) & Avg leverage \\
\midrule
BTC  & Binance funding      & 9.73 & 9.70 & 10.33 & -0.06 & 4.41 \\
BTC  & Hyperliquid funding  & 20.83 & 20.77 & 21.39 & -0.07 & 3.89 \\
DOGE & Binance funding      & 7.68 & 7.66 & 9.49 & -0.26 & 1.85 \\
DOGE & Hyperliquid funding  & 25.32 & 25.24 & 26.98 & -0.24 & 1.66 \\
ETH  & Binance funding      & 10.06 & 10.03 & 10.83 & -0.14 & 3.95 \\
ETH  & Hyperliquid funding  & 18.31 & 18.25 & 18.99 & -0.18 & 3.58 \\
LINK & Binance funding      & 8.28 & 8.26 & 9.58 & -0.17 & 2.15 \\
LINK & Hyperliquid funding  & 18.95 & 18.89 & 20.30 & -0.16 & 1.92 \\
\bottomrule
\end{tabular}
\end{table*}

\begin{table*}[t]
\centering
\caption{Historical rebalancing diagnostics under alternative funding environments, scaled by the initial AuM}
\label{tab:backtest-rebalance-diagnostics}
\small
\setlength{\tabcolsep}{6pt}
\begin{tabular}{llccc}
\toprule
Ticker & Funding environment & Rebalances & Turnover (\% of initial AuM) & Avg rebalancing (\% of initial AuM) \\
\midrule
BTC  & Binance funding      & 5 & 67.86\%  & 13.57\% \\
BTC  & Hyperliquid funding  & 5 & 71.37\%  & 14.27\% \\
DOGE & Binance funding      & 10 & 232.78\% & 23.28\% \\
DOGE & Hyperliquid funding  & 9 & 224.65\% & 24.96\% \\
ETH  & Binance funding      & 7 & 111.78\% & 15.97\% \\
ETH  & Hyperliquid funding  & 6 & 102.39\% & 17.06\% \\
LINK & Binance funding      & 6 & 140.31\% & 23.39\% \\
LINK & Hyperliquid funding  & 6 & 150.96\% & 25.16\% \\
\bottomrule
\end{tabular}
\end{table*}

Together, these results answer the most important empirical question for the backtest layer. The historically calibrated policy is not merely internally coherent: it remains operationally stable out of time, agrees directionally with earlier leverage-based research, and produces materially different realized economics when the funding environment changes. The funding claim is therefore not a reduced form narrative overlay but an observed out-of-time property of the same frozen control rule.

\section{Execution Validation}

\subsection{Live Deployment Setup}

To validate the economic relevance of the execution layer, we study a live basis-trading deployment implemented as an ERC-4626 tokenized vault on Arbitrum \cite{ERC4626}. The strategy was active from 2025-04-01 through 2025-12-01. The spot leg was executed through 1inch's routed quote and RFQ-capable infrastructure rather than by interacting with a single AMM pool directly. This choice is consistent with the recent solver-based execution literature, which finds that off-chain competition and routed liquidity access can improve execution welfare relative to vanilla AMM routing, and with the 1inch API stack, which exposes both routed swap endpoints and orderbook-based RFQ functionality on EVM chains including Arbitrum \cite{YuminagaChenSuiExecutionWelfare,OneInchClassicSwapAPI,OneInchOrderbookAPI}. The short hedge was executed on Hyperliquid's fully on-chain perpetual order book \cite{HyperliquidOrderBook,HyperliquidInfoEndpoint}.

Table~\ref{tab:execution-addresses} records the public addresses corresponding to the validation deployment.

\begin{table}[t]
\centering
\caption{Public addresses for the live execution validation deployment.}
\label{tab:execution-addresses}
\begin{tabular}{ll}
\toprule
Component & Address \\
\midrule
Vault & \href{https://arbiscan.io/address/0xe5fc579f20C2dbffd78a92ddD124871a35519659}{0xe5fc579f20C2dbffd78a92ddD124871a35519659} \\
Hyperliquid & \href{https://hypurrscan.io/address/0xEdD0D94a267550C926cbe4A1A562157945cc6c6c}{0xEdD0D94a267550C926cbe4A1A562157945cc6c6c} \\
Spot & \href{https://arbiscan.io/address/0x28D21b1B23440DEc140D74f569a0Aeb98B0C5201}{0x28D21b1B23440DEc140D74f569a0Aeb98B0C5201} \\
\bottomrule
\end{tabular}
\end{table}

\subsection{Expected Versus Realized Spread Diagnostics}

We analyze all basis trades executed over the live sample and compare the contemporaneous expected execution cost with the realized execution cost. Let $K_i^{\mathrm{target}}$ denote the controller's contemporaneous basis-trade cost target for trade $i$, measured from a rolling three-second average built from 1inch spot quotes and executable prices in Hyperliquid's perpetual order book \cite{OneInchClassicSwapAPI,HyperliquidOrderBook}. Let $C_i^{\mathrm{real}}$ denote the realized effective execution cost. Operationally, the spot leg was always executed first, and the perpetual hedge was executed second. This sequencing is economically important: the realized execution cost therefore already embeds routed spot slippage, impact, and quote-to-fill drift, even though the summary statistics reported below exclude blockchain transaction costs and venue fees.

\begin{figure}[t]
    \centering
    \includegraphics[width=\columnwidth]{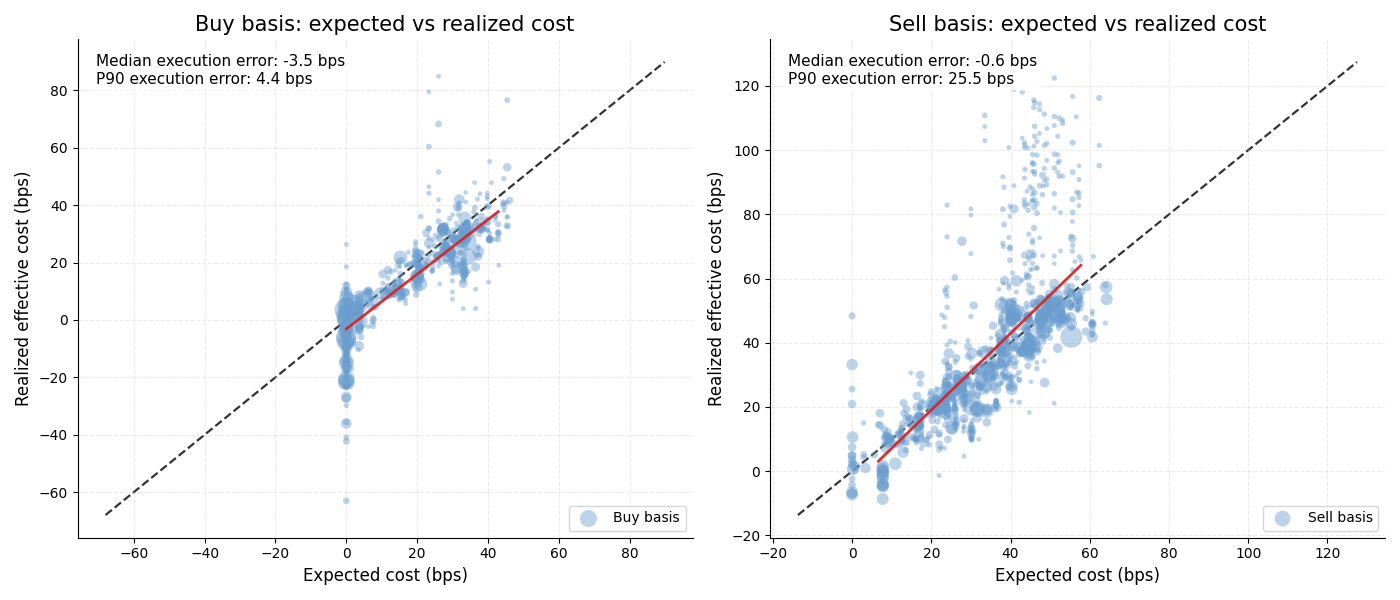}
    \caption{Expected versus realized effective execution cost by side. Point size is proportional to trade notional. The dashed line is the $45^\circ$ reference line, while the solid fitted line summarizes the empirical expected--realized mapping on each side.}
    \label{fig:expected-realized-live}
\end{figure}

The live validation sample contains 1,924 basis trades and approximately \$8.95 million of aggregate spot notional. Table~\ref{tab:execution-summary-side} reports the basic cross-sectional split of the execution sample. Figure~\ref{fig:expected-realized-live} then compares $K_i^{\mathrm{target}}$ with $C_i^{\mathrm{real}}$ separately for buy-basis and sell-basis trades. The expected execution signal is informative on both sides: realized costs co-move strongly with the contemporaneous target. At the same time, the sell-basis side is visibly noisier and exhibits a materially thicker right tail of execution errors. In the live sample, the median execution error $C_i^{\mathrm{real}}-K_i^{\mathrm{target}}$ is modestly negative on both sides, but the 90th percentile is much higher for the sell-basis trades. This pattern supports a side-specific implementation rule in which execution asymmetry is absorbed through side-specific cost budgets rather than through a common rebalancing threshold.

\begin{table}[t]
\centering
\caption{Execution volume and median spread for the live validation sample. Reported spreads already include routed slippage, impact, and spread, but exclude blockchain transaction costs and venue fees.}
\label{tab:execution-summary-side}
\small
\setlength{\tabcolsep}{4pt}
\begin{tabular}{lccc}
\toprule
Side & Trades & Volume (USD mm) & Median $|\mathrm{spread}|$ (bps) \\
\midrule
Buy basis & 675 & 4.35 & 19.0 \\
Sell basis & 1249 & 4.59 & 38.9 \\
All trades & 1924 & 8.95 & 29.1 \\
\bottomrule
\end{tabular}
\end{table}

Four findings stand out. First, realized execution wedges are economically material: the median absolute spread is about 29 bps and remains close to 39 bps on the sell-basis side. Second, the side asymmetry is large and stable: buy-basis trades exhibit a median effective cost of about 18 bps, whereas sell-basis trades are closer to 39 bps, with correspondingly larger upper-tail costs. Hence,
\begin{equation}
    K_{\mathrm{reb}}^{\mathrm{sell\ basis}} > K_{\mathrm{reb}}^{\mathrm{buy\ basis}},
\end{equation}
which directly supports the asymmetry of the dynamic control rules. Third, the expected execution signal is useful but imperfect: the buy-basis side is moderately conservative in the median, while the sell-basis side is approximately unbiased in the median but exhibits a significantly larger right-tail execution error. Fourth, bootstrap confidence intervals around the main cost statistics are sufficiently tight to make the side asymmetry economically and statistically meaningful: for example, the raw sample median effective cost is about 18.2 bps for buy basis (95\% bootstrap CI roughly 15.9--19.1 bps) versus 38.9 bps for sell basis (95\% bootstrap CI roughly 38.3--40.0 bps).

\subsection{Trade-Level Win Rate, Minimum Size, and Buffer Calibration}

We next convert the live execution sample into a practical calibration exercise for the intervention rule. Let $K_i^{\mathrm{target}}$ denote the contemporaneous target cost inferred from the spot--perpetual quote pair used by the controller. Because the vault is implemented as an ERC-4626 vehicle, deposits and withdrawals mint or burn shares immediately at the prevailing state, which is economically equivalent in our setup to an immediate user-side fixation of the basis spread. The relevant target is therefore time-varying rather than constant. For a candidate execution buffer $b \ge 0$, define the trade level success indicator, the ordinary win rate, and the capital weighted win rate as
\begin{align}
    W_i(b)
    &:= \mathbf{1}\!\left\{C_i^{\mathrm{real}} \le K_i^{\mathrm{target}} + b\right\},
    \label{eq:win-indicator}
    \\
    \mathrm{WR}(b)
    &:= \frac{1}{N}\sum_{i=1}^N W_i(b),
    \label{eq:win-rate}
    \\
    \mathrm{WR}^{\$}(b)
    &:= \frac{\sum_{i=1}^N q_i W_i(b)}{\sum_{i=1}^N q_i},
    \label{eq:cap-weighted-win-rate}
\end{align}
where $q_i$ is the executed notional.

\begin{figure}[t]
    \centering
    \includegraphics[width=\columnwidth]{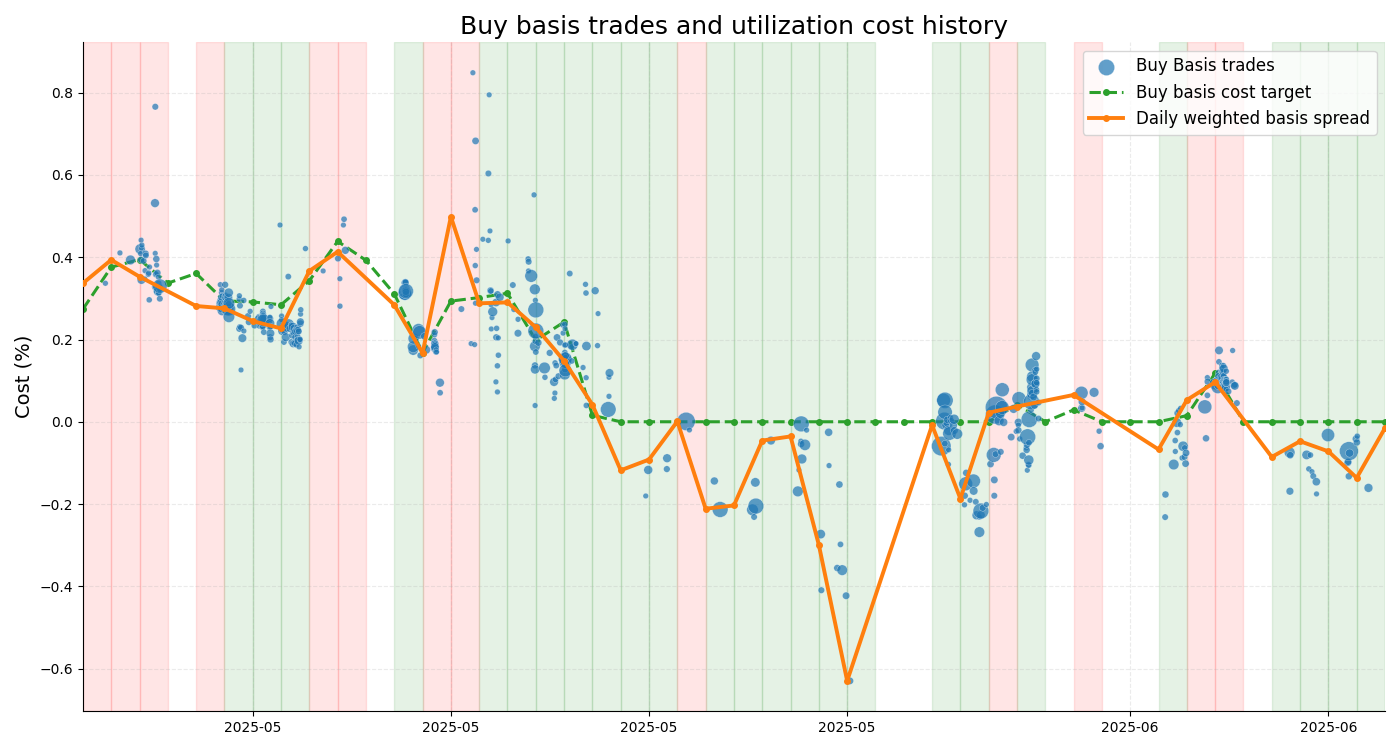}
    \caption{Buy-basis trade history and floating buy-basis target. The spot leg is executed first, so the realized cost already reflects routed slippage, impact, and quote-to-fill drift.}
    \label{fig:buy-history-live}
\end{figure}

\begin{figure}[t]
    \centering
    \includegraphics[width=\columnwidth]{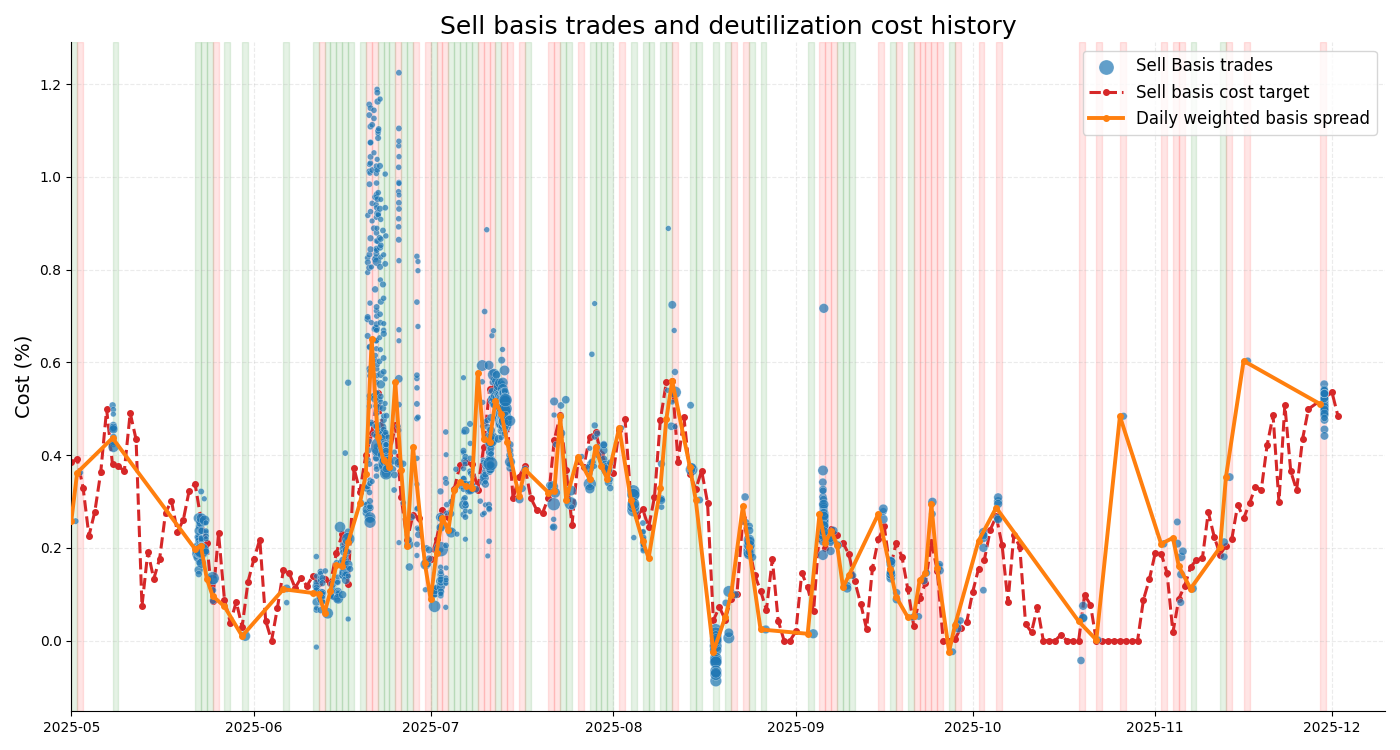}
    \caption{Sell-basis trade history and floating sell-basis target. The realized wedge is materially more dispersed than on the buy-basis side, especially during wider-spread market states.}
    \label{fig:sell-history-live}
\end{figure}

Figures~\ref{fig:buy-history-live} and \ref{fig:sell-history-live} report the trade histories of the buy-basis and the sell-basis. They show that the target is genuinely floating in live operation because the controller updates the admissible execution budget together with the vault state and the contemporaneous spread environment. The figures also make clear that the execution problem is side-asymmetric and size-dependent: sell-basis trades operate against a noisier and more costly microstructure, while very small trades are disproportionately exposed to short horizon spread volatility. This motivates two practical implementation parameters. The first is a minimum trade size. Based on the observed fixed transaction-cost layer and the fixed Hyperliquid withdrawal charge, we adopt an empirical lower bound of approximately \$10k for economically meaningful rebalances. The second is the execution buffer $b$, which absorbs the quote-to-fill execution error and also acts as protection against misspecification in the effective funding gain available at the moment of rebalancing.

\begin{table}[t]
\centering
\caption{Raw versus post-filter execution summary. The post-filter sample retains only trades with executed notional of at least \$10k.}
\label{tab:execution-prefilter-postfilter}
\small
\setlength{\tabcolsep}{4pt}
\begin{tabular}{llcccc}
\toprule
Sample & Side & Trades & Median (bps) & Win rate & Weighted WR \\
\midrule
Raw & All & 1924 & 28.9 & 87.9\% & 95.0\% \\
Raw & Buy basis & 675 & 18.2 & 95.6\% & 95.4\% \\
Raw & Sell basis & 1249 & 38.9 & 83.7\% & 94.6\% \\
\midrule
Post-filter & All & 209 & 22.1 & 93.8\% & 95.2\% \\
Post-filter & Buy basis & 78 & 8.8 & 93.6\% & 95.4\% \\
Post-filter & Sell basis & 131 & 31.5 & 93.9\% & 94.9\% \\
\bottomrule
\end{tabular}
\end{table}

\begin{figure}[t]
    \centering
    \includegraphics[width=\columnwidth]{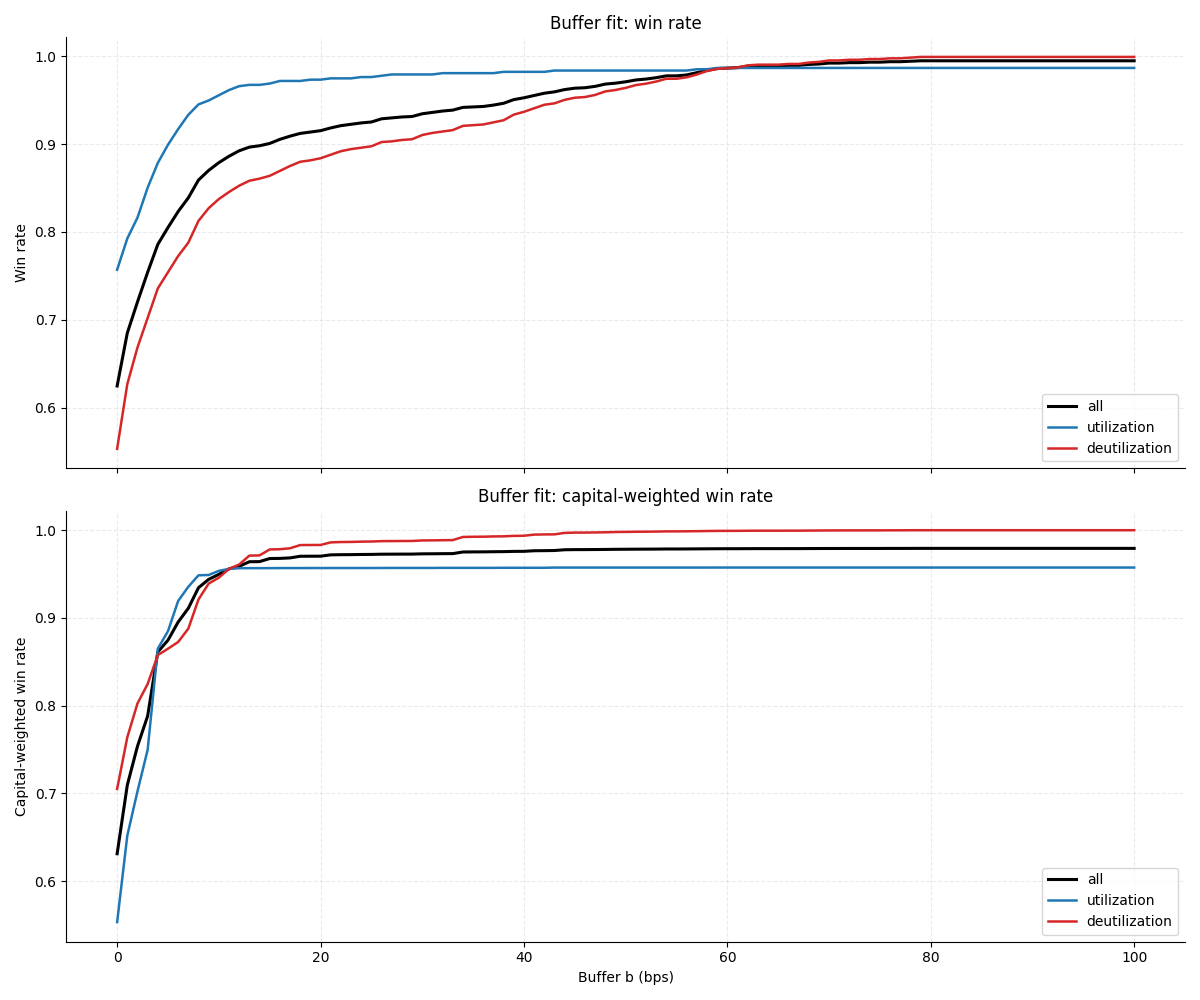}
    \caption{Trade-level and capital-weighted success as functions of the execution buffer. The buffer is calibrated against the floating live target rather than against a fixed ex ante threshold.}
    \label{fig:buffer-fit-live}
\end{figure}

The filter and buffer exercise yields three practical conclusions. First, the minimum-size filter is economically meaningful: after restricting attention to trades above \$10k, the median effective cost falls from about 28.9 bps to 22.1 bps in the aggregate sample, with the buy-basis side especially improving sharply. Second, capital-weighted execution quality is materially stronger than raw trade-count success. Even in the unfiltered sample, the capital-weighted win rate is already about 95\%, indicating that the economically relevant trades execute much better than the small-trade tail of the distribution. Third, the execution buffer required for robust live operation is modest but strictly positive. In the post-filter sample, a buffer on the order of 10 bps is sufficient to reach approximately 95\% capital-weighted success, which we interpret as a practical calibration target for live deployment.

The live validation therefore sharpens the practical interpretation of the dynamic layer. It confirms that execution costs are neither symmetric across directions nor constant across sizes; that the sell-basis leg is materially more expensive than the buy-basis leg; that large trades can be executed more efficiently than small ones under routed spot access; and that a modest positive buffer is necessary even after one uses a dynamic target. In short, execution constraints are side-asymmetric and size-dependent. This supports the final policy prescription of the paper: a benchmark target inherited from the static constrained problem, a lower liquidation-budget limit, a side-aware and size-aware execution filter, an empirical minimum size of roughly \$10k, and an execution buffer on the order of 10 bps for live deployment.

\section{Liquidity Constraints and Strategy Capacity}

The dynamic control problem is incomplete without an implementation layer. A theoretical trigger at $\alpha_t<\alpha_L$ or $\alpha_t>\alpha_U$ is operationally meaningful only if the required basis trade can actually be executed at an acceptable cost and in an economically significant size. We therefore impose side-specific liquidity constraints that map instantaneous execution conditions into admissible rebalancing sizes and, in turn, into a strategy-level capacity object. This construction is consistent with the standard execution literature such as \cite{AlmgrenChriss2001,Kyle1985,ObizhaevaWang2013,Hasbrouck2007}, in which market impact and depth constraints are modeled jointly with optimal trading decisions rather than appended ex post.

\subsection{Instantaneous and Operational Basis-Trade Costs}

We begin with the contemporaneous executable basis-trade costs introduced in Section~III. Let
\begin{equation}
    C^{\mathrm{inst}}_{\mathrm{buy}}(q,t)
\end{equation}
denote the cost of executing a buy-basis trade of USD notional $q$ at time $t$, i.e., buying a spot and selling a perpetual hedge. Let
\begin{equation}
    C^{\mathrm{inst}}_{\mathrm{sell}}(q,t)
\end{equation}
denote the cost of executing a sell-basis trade, i.e., selling spot and buying back the perpetual hedge. For analytical clarity, we decompose these objects into leg level costs:
\begin{align}
    C^{\mathrm{inst}}_{\mathrm{buy}}(q,t)
    &=
    C^{\mathrm{inst}}_{\mathrm{spot},+}(q,t)
    +
    C^{\mathrm{inst}}_{\mathrm{perp},-}(q,t)
    +
    F_{\mathrm{buy}}(t),
    \label{eq:c-buy-inst}
    \\
    C^{\mathrm{inst}}_{\mathrm{sell}}(q,t)
    &=
    C^{\mathrm{inst}}_{\mathrm{spot},-}(q,t)
    +
    C^{\mathrm{inst}}_{\mathrm{perp},+}(q,t)
    +
    F_{\mathrm{sell}}(t),
    \label{eq:c-sell-inst}
\end{align}
where $C^{\mathrm{inst}}_{\mathrm{spot},+}$ and $C^{\mathrm{inst}}_{\mathrm{spot},-}$ are the currently executable spot-leg costs for buying and selling spot, $C^{\mathrm{inst}}_{\mathrm{perp},+}$ and $C^{\mathrm{inst}}_{\mathrm{perp},-}$ are the corresponding perpetual-leg costs, and $F_{\mathrm{buy}}(t)$, $F_{\mathrm{sell}}(t)$ collect fixed frictions such as gas, venue fees, and withdrawal or bridge overhead. The live validation in Section~X shows that these objects are directionally asymmetric, so, in general,
\begin{equation}
    C^{\mathrm{inst}}_{\mathrm{buy}}(q,t)\neq C^{\mathrm{inst}}_{\mathrm{sell}}(q,t).
\end{equation}

Instantaneous costs are the correct objects for execution diagnostics, but they are too noisy to serve directly as liquidity constraints at the strategy level. We therefore work with operational cost functions obtained from recent snapshots or fills:
\begin{equation}
    \widehat C_s(q,t)
    =
    \mathcal F\!\left(
        \{C^{\mathrm{inst}}_s(q,\tau)\}_{\tau\in[t-\Delta,t]}
    \right),
    \qquad
    s\in\{\mathrm{buy},\mathrm{sell}\},
    \label{eq:c-hat}
\end{equation}
where $\mathcal F$ is a recent-history estimator such as a rolling median, a stressed quantile, or a simple parametric fit. In the empirical implementation, the 1inch RFQ and routing sample provide the spot-side inputs, while Hyperliquid L2 snapshots and fills provide the perpetual-side inputs.

\subsection{Admissible Rebalancing Size}

Let $K_{\mathrm{buy}}^{\max}(t)$ and $K_{\mathrm{sell}}^{\max}(t)$ denote side-specific cost budgets. The corresponding instantaneous executable sizes are
\begin{align}
    Q_{\mathrm{buy}}^{\max}(t)
    &:=
    \sup\{q\ge 0:\ \widehat C_{\mathrm{buy}}(q,t)\le K_{\mathrm{buy}}^{\max}(t)\},
    \label{eq:qmax-buy}
    \\
    Q_{\mathrm{sell}}^{\max}(t)
    &:=
    \sup\{q\ge 0:\ \widehat C_{\mathrm{sell}}(q,t)\le K_{\mathrm{sell}}^{\max}(t)\}.
    \label{eq:qmax-sell}
\end{align}
Equations~\eqref{eq:qmax-buy}--\eqref{eq:qmax-sell} can also be interpreted leg by leg by replacing the total cost budget with spot-side and perpetual-side budgets and then taking the minimum of the resulting leg-level capacities. For the present paper, the combined basis-trade formulation is sufficient and keeps the notation aligned with the execution sample.

The second restriction is economic rather than microstructural. Small rebalances are dominated by fixed frictions and should not be executed even if the book is locally deep. We therefore define minimum economically meaningful sizes
\begin{equation}
    q_{\min,\mathrm{buy}}>0,
    \qquad
    q_{\min,\mathrm{sell}}>0,
\end{equation}
which in the live deployment are summarized by the empirical threshold of roughly \$10k reported in Section~X. The admissible rebalancing sets are then
\begin{equation}
    \mathcal Q_{\mathrm{buy}}(t)
    =
    [q_{\min,\mathrm{buy}},\,Q_{\mathrm{buy}}^{\max}(t)],
    \mathcal Q_{\mathrm{sell}}(t)
    =
    [q_{\min,\mathrm{sell}},\,Q_{\mathrm{sell}}^{\max}(t)].
    \label{eq:q-admissible}
\end{equation}
If $Q_s^{\max}(t)<q_{\min,s}$ for $s\in\{\mathrm{buy},\mathrm{sell}\}$, then no economically significant rebalancing is executable on the side $s$ at time $t$.

\subsection{Rebalancing Feasibility and Strategy Capacity}

Let $q_{\mathrm{buy}}^{\ast}(t)$ and $q_{\mathrm{sell}}^{\ast}(t)$ denote the trade sizes implied by the dynamic control rule when the strategy attempts to move from the realized state $\alpha_t$ back toward the benchmark target $\alpha^{\dagger}$. A lower-side intervention is operationally feasible only if
\begin{equation}
    \alpha_t<\alpha_L
    \quad\text{and}\quad
    q_{\mathrm{buy}}^{\ast}(t)\in\mathcal Q_{\mathrm{buy}}(t),
    \label{eq:liq-lower-feas}
\end{equation}
whereas an upper-side intervention is feasible only if
\begin{equation}
    \alpha_t>\alpha_U
    \quad\text{and}\quad
    q_{\mathrm{sell}}^{\ast}(t)\in\mathcal Q_{\mathrm{sell}}(t).
    \label{eq:liq-upper-feas}
\end{equation}
Hence, a control trigger is operationally valid only if the required rebalancing size lies inside the corresponding admissible liquidity set.

To translate this condition into strategy capacity, write the required rebalancing size on the side $s$ as
\begin{equation}
    q_s^{\ast}(t)
    =
    \phi_s(\alpha_t,\alpha^{\dagger})V_t,
    \qquad
    s\in\{\mathrm{buy},\mathrm{sell}\},
    \label{eq:phi-capacity}
\end{equation}
where $V_t$ is the equity in the current vault and $\phi_s$ is the side-specific fraction of capital that must be reallocated to restore the target. The corresponding side-specific capacity is
\begin{equation}
    V_s^{\max}(t)
    =
    \frac{Q_s^{\max}(t)}{\phi_s(\alpha_t,\alpha^{\dagger})},
    \qquad
    s\in\{\mathrm{buy},\mathrm{sell}\},
    \label{eq:side-capacity}
\end{equation}
and the whole-strategy instantaneous capacity is
\begin{equation}
    V^{\max}(t)
    =
    \min\bigl(V_{\mathrm{buy}}^{\max}(t),\,V_{\mathrm{sell}}^{\max}(t)\bigr).
    \label{eq:whole-capacity}
\end{equation}
If a single robust scalar is required, the natural summary is a quantile capacity,
\begin{equation}
    V_{\mathrm{safe}}^{\max}
    :=
    \operatorname{Quantile}_{p}\!\left(V^{\max}(t)\right),
    \qquad p\in(0,1),
    \label{eq:safe-capacity}
\end{equation}
for example, with $p=0.25$ as a conservative capacity benchmark.

The economic implication is immediate. Liquidity constraints are side-dependent through $C_{\mathrm{buy}}$ versus $C_{\mathrm{sell}}$, size-dependent through the mapping $q\mapsto \widehat C_s(q,t)$, and state-dependent because the required rebalancing fraction $\phi_s(\alpha_t,\alpha^{\dagger})$ depends on the realized control state. A theoretically optimal intervention is therefore not sufficient: the basis trade is implementable only when the contemporaneous liquidity environment supports the required rebalancing at admissible cost and scale.

\subsection{Limitations and Next Steps}

Three limitations define the natural extensions of the paper. First, the liquidation-risk layer is built on a tractable diffusion benchmark rather than on a fully structural model of crypto tail risk. This is a deliberate control-design choice, partially mitigated in practice through stressed volatility, conservative liquidation horizons, strict liquidation budgets, and the bootstrap robustness exercise for the lower boundary. Second, the Binance--Hyperliquid comparison is in a reduced form rather than causal: the observed gap in benchmark collateral shares reflects the joint effect of margin architecture, realized volatility, and funding conditions. A natural next step is to decompose these channels more explicitly using richer venue-level microstructure data. Third, the paper formalizes liquidity constraints analytically, but leaves a full empirical capacity frontier based on synchronized RFQ and perpetual-order-book data for future work.

More broadly, the next research step is to deepen the microstructure layer for perpetual DEX markets. On the control side, a natural extension is to replace the current dynamic rule-based layer with a continuous time impulse control formulation in the spirit of the classical intervention literature such as \cite{OksendalSulem2007}.

\section{Conclusion}

This paper formulates permissionless spot--perpetual basis trading as a collateral-control problem. The central object is the collateral share posted to the derivative leg, which turns the strategy from a leverage heuristic into an explicit allocation-and-intervention rule.

The static result is that the risk-constrained formulation provides the more useful operating benchmark relative to the economic optimum. The dynamic result is that the control problem is asymmetric: the lower trigger is solvency-driven, whereas the upper trigger is economic and survives mainly in favorable carry-cost regimes. The implementation result is that execution frictions are first order: live routed execution shows material side asymmetry and a meaningful minimum rebalancing size, while the historical backtest layer shows that, holding the control rule fixed, realized performance is predominantly shaped by the funding environment.

Together, these results imply that permissionless basis trading requires explicit collateral rules that jointly account for solvency, execution, and liquidity constraints. More broadly, the paper contributes a static benchmark control, an asymmetric dynamic extension, and an execution-aware validation framework for permissionless spot--perpetual basis trading.

\section{Data and Code Availability}

The Jupyter notebooks, processed datasets, and
scripts reproducing the comparative optimal control analysis
(Section~VI), the bootstrap robustness diagnostics (Section~VII), the
funding environment backtest visualizations (Section~IX), and the
execution diagnostics (Section~X) are publicly available
at~\cite{SimulationsGitHub}. The repository includes the live BasisOS
trade extracts, the daily vault entry and exit costs time series, and the
per-ticker basis trade backtests outputs consumed by the
funding diagnostics. The historical 2023--2024 calibration
code referenced in Section~IX and Appendix~D is distributed separately
as part of our DeFi research framework~\cite{FractalDeFiStrategy}.

Raw market data are sourced from the public Binance REST endpoints
(spot, mark, and funding) and from the Hyperliquid Info API; both are
free and unauthenticated, and the relevant pull scripts are included
in the repository. The live execution sample (Section~X) is
reconstructed from on-chain transactions of the public vault
addresses listed in Table~\ref{tab:execution-addresses}; raw
transaction logs are accessible via Arbiscan and Hypurrscan, and
the parsing scripts producing the trade-level diagnostics are
included in the same repository. No proprietary datasets, vendor
feeds, or restricted credentials are required to reproduce any
result reported in the paper.

\clearpage
\bibliography{bibliography}

\clearpage
\onecolumn
\appendices

\section{Extended Static Tables}
Tables~\ref{tab:btc-grid}--\ref{tab:doge-grid} report the extended static simulation grid by asset.\vspace{-0.4em}

\begin{table}[H]
\centering
\caption{BTC: static grid. Panel A reports the economic formulation with $LGD=0.1$; Panel B reports the risk-constrained formulation with $\varepsilon=0.001$. For Binance, $\theta_F=1/L_{\max}$; for Hyperliquid, $\theta_F=1/(2L_{\max})$.}
\label{tab:btc-grid}
\small
\begin{tabular}{llcc|ccc|ccc}
\toprule
Venue & Lookback & $L_{\max}$ & $\theta_F$ & \multicolumn{3}{c|}{Panel A: Economic formulation} & \multicolumn{3}{c}{Panel B: Risk-constrained formulation}\\
 &  &  &  & $1.0\times$ & $1.5\times$ & $2.0\times$ & $1.0\times$ & $1.5\times$ & $2.0\times$ \\
\midrule
Binance & 30d  & 125 & 0.0080 & 0.127 & 0.172 & 0.211 & 0.092 & 0.131 & 0.167 \\
Binance & 90d  & 125 & 0.0080 & 0.126 & 0.169 & 0.205 & 0.088 & 0.128 & 0.164 \\
Binance & 180d & 125 & 0.0080 & 0.123 & 0.157 & 0.192 & 0.086 & 0.123 & 0.158 \\
Binance & 360d & 125 & 0.0080 & 0.111 & 0.141 & 0.172 & 0.079 & 0.111 & 0.143 \\
Hyperliquid & 30d  & 40 & 0.0125 & 0.131 & 0.178 & 0.220 & 0.095 & 0.135 & 0.171 \\
Hyperliquid & 90d  & 40 & 0.0125 & 0.128 & 0.173 & 0.213 & 0.093 & 0.132 & 0.169 \\
Hyperliquid & 180d & 40 & 0.0125 & 0.122 & 0.157 & 0.193 & 0.089 & 0.127 & 0.163 \\
Hyperliquid & 360d & 40 & 0.0125 & 0.119 & 0.150 & 0.180 & 0.088 & 0.124 & 0.158 \\
\bottomrule
\end{tabular}
\end{table}

\begin{table}[H]
\centering
\caption{ETH: static grid. Panel A reports the economic formulation with $LGD=0.1$; Panel B reports the risk-constrained formulation with $\varepsilon=0.001$.}
\label{tab:eth-grid}
\small
\begin{tabular}{llcc|ccc|ccc}
\toprule
Venue & Lookback & $L_{\max}$ & $\theta_F$ & \multicolumn{3}{c|}{Panel A: Economic formulation} & \multicolumn{3}{c}{Panel B: Risk-constrained formulation}\\
 &  &  &  & $1.0\times$ & $1.5\times$ & $2.0\times$ & $1.0\times$ & $1.5\times$ & $2.0\times$ \\
\midrule
Binance & 30d  & 100 & 0.0100 & 0.170 & 0.221 & 0.265 & 0.117 & 0.164 & 0.209 \\
Binance & 90d  & 100 & 0.0100 & 0.177 & 0.232 & 0.280 & 0.115 & 0.164 & 0.210 \\
Binance & 180d & 100 & 0.0100 & 0.198 & 0.267 & 0.327 & 0.117 & 0.167 & 0.214 \\
Binance & 360d & 100 & 0.0100 & 0.204 & 0.275 & 0.337 & 0.120 & 0.170 & 0.217 \\
Hyperliquid & 30d  & 25 & 0.0200 & 0.165 & 0.214 & 0.257 & 0.125 & 0.172 & 0.217 \\
Hyperliquid & 90d  & 25 & 0.0200 & 0.186 & 0.248 & 0.303 & 0.124 & 0.173 & 0.219 \\
Hyperliquid & 180d & 25 & 0.0200 & 0.195 & 0.262 & 0.320 & 0.126 & 0.176 & 0.223 \\
Hyperliquid & 360d & 25 & 0.0200 & 0.204 & 0.275 & 0.337 & 0.126 & 0.174 & 0.219 \\
\bottomrule
\end{tabular}
\end{table}

\begin{table}[H]
\centering
\caption{LINK: static grid. Panel A reports the economic formulation with $LGD=0.1$; Panel B reports the risk-constrained formulation with $\varepsilon=0.001$.}
\label{tab:link-grid}
\small
\begin{tabular}{llcc|ccc|ccc}
\toprule
Venue & Lookback & $L_{\max}$ & $\theta_F$ & \multicolumn{3}{c|}{Panel A: Economic formulation} & \multicolumn{3}{c}{Panel B: Risk-constrained formulation}\\
 &  &  &  & $1.0\times$ & $1.5\times$ & $2.0\times$ & $1.0\times$ & $1.5\times$ & $2.0\times$ \\
\midrule
Binance & 30d  & 50 & 0.0200 & 0.192 & 0.254 & 0.308 & 0.121 & 0.166 & 0.209 \\
Binance & 90d  & 50 & 0.0200 & 0.216 & 0.287 & 0.347 & 0.125 & 0.175 & 0.221 \\
Binance & 180d & 50 & 0.0200 & 0.246 & 0.328 & 0.396 & 0.143 & 0.200 & 0.253 \\
Binance & 360d & 50 & 0.0200 & 0.253 & 0.336 & 0.405 & 0.148 & 0.205 & 0.259 \\
Hyperliquid & 30d  & 10 & 0.0500 & 0.200 & 0.265 & 0.320 & 0.144 & 0.190 & 0.232 \\
Hyperliquid & 90d  & 10 & 0.0500 & 0.226 & 0.298 & 0.360 & 0.150 & 0.198 & 0.243 \\
Hyperliquid & 180d & 10 & 0.0500 & 0.258 & 0.338 & 0.408 & 0.167 & 0.223 & 0.274 \\
Hyperliquid & 360d & 10 & 0.0500 & 0.263 & 0.345 & 0.414 & 0.167 & 0.221 & 0.271 \\
\bottomrule
\end{tabular}
\end{table}

\begin{table}[H]
\centering
\caption{DOGE: static grid. Panel A reports the economic formulation with $LGD=0.1$; Panel B reports the risk-constrained formulation with $\varepsilon=0.001$.}
\label{tab:doge-grid}
\small
\begin{tabular}{llcc|ccc|ccc}
\toprule
Venue & Lookback & $L_{\max}$ & $\theta_F$ & \multicolumn{3}{c|}{Panel A: Economic formulation} & \multicolumn{3}{c}{Panel B: Risk-constrained formulation}\\
 &  &  &  & $1.0\times$ & $1.5\times$ & $2.0\times$ & $1.0\times$ & $1.5\times$ & $2.0\times$ \\
\midrule
Binance & 30d  & 50 & 0.0200 & 0.213 & 0.290 & 0.356 & 0.129 & 0.178 & 0.223 \\
Binance & 90d  & 50 & 0.0200 & 0.235 & 0.313 & 0.381 & 0.135 & 0.188 & 0.238 \\
Binance & 180d & 50 & 0.0200 & 0.260 & 0.340 & 0.409 & 0.143 & 0.201 & 0.254 \\
Binance & 360d & 50 & 0.0200 & 0.268 & 0.349 & 0.418 & 0.150 & 0.209 & 0.263 \\
Hyperliquid & 30d  & 10 & 0.0500 & 0.225 & 0.304 & 0.372 & 0.152 & 0.200 & 0.244 \\
Hyperliquid & 90d  & 10 & 0.0500 & 0.249 & 0.327 & 0.393 & 0.159 & 0.211 & 0.259 \\
Hyperliquid & 180d & 10 & 0.0500 & 0.273 & 0.352 & 0.419 & 0.168 & 0.224 & 0.276 \\
Hyperliquid & 360d & 10 & 0.0500 & 0.280 & 0.359 & 0.426 & 0.170 & 0.225 & 0.276 \\
\bottomrule
\end{tabular}
\end{table}
\clearpage
\section{Monte Carlo Upper-Bound Evidence}
Table~\ref{tab:mc-upper} summarizes the Monte Carlo upper-bound exercise with benchmark targets fixed from Table~\ref{tab:main} and expected funding evaluated over a 14-day period.

\begin{table}[H]
\centering
\caption{Monte Carlo upper-bound summary. The target $\alpha^{\dagger}$ is fixed at the benchmark constrained optimum, expected funding is measured over 14 days, and the rebalance-cost scenarios are 5, 10, 20, and 30 bps of capital.}
\label{tab:mc-upper}
\small
\begin{tabular}{llcccccc}
\toprule
Venue & Asset & $\alpha^{\dagger}$ & Cost & $\widetilde{\kappa}_{14d}$ & $\alpha_U$ & Hit rate & Median hit (d) \\
\midrule
Binance & BTC  & 0.123 & 5  & 11.31 bps & 0.565 & 0.090 & 48.4 \\
Binance & ETH  & 0.167 & 5  & 8.73 bps  & 0.740 & 0.035 & 52.1 \\
Binance & LINK & 0.200 & 5  & 15.38 bps & 0.525 & 0.592 & 29.6 \\
Binance & DOGE & 0.201 & 5  & 3.86 bps  & 0.990 & 0.000 & -- \\
Hyperliquid & BTC  & 0.127 & 5  & 22.91 bps & 0.345 & 0.599 & 27.9 \\
Hyperliquid & ETH  & 0.176 & 5  & 23.78 bps & 0.386 & 0.735 & 20.6 \\
Hyperliquid & LINK & 0.223 & 5  & 36.34 bps & 0.361 & 0.890 & 10.2 \\
Hyperliquid & DOGE & 0.224 & 5  & 7.20 bps  & 0.918 & 0.000 & -- \\
\midrule
Binance & BTC  & 0.123 & 10 & 11.31 bps & 0.990 & 0.000 & -- \\
Binance & ETH  & 0.167 & 10 & 8.73 bps  & 0.990 & 0.000 & -- \\
Binance & LINK & 0.200 & 10 & 15.38 bps & 0.850 & 0.011 & 54.9 \\
Binance & DOGE & 0.201 & 10 & 3.86 bps  & 0.990 & 0.000 & -- \\
Hyperliquid & BTC  & 0.127 & 10 & 22.91 bps & 0.564 & 0.090 & 48.5 \\
Hyperliquid & ETH  & 0.176 & 10 & 23.78 bps & 0.597 & 0.267 & 42.4 \\
Hyperliquid & LINK & 0.223 & 10 & 36.34 bps & 0.498 & 0.670 & 25.6 \\
Hyperliquid & DOGE & 0.224 & 10 & 7.20 bps  & 0.990 & 0.000 & -- \\
\bottomrule
\end{tabular}
\end{table}

\clearpage
\section{Raw Funding Diagnostics}
\begin{figure}[H]
    \centering
    \setlength{\abovecaptionskip}{4pt}
    \setlength{\belowcaptionskip}{0pt}
    \begin{subfigure}[t]{0.495\textwidth}
        \centering
        \includegraphics[width=\linewidth]{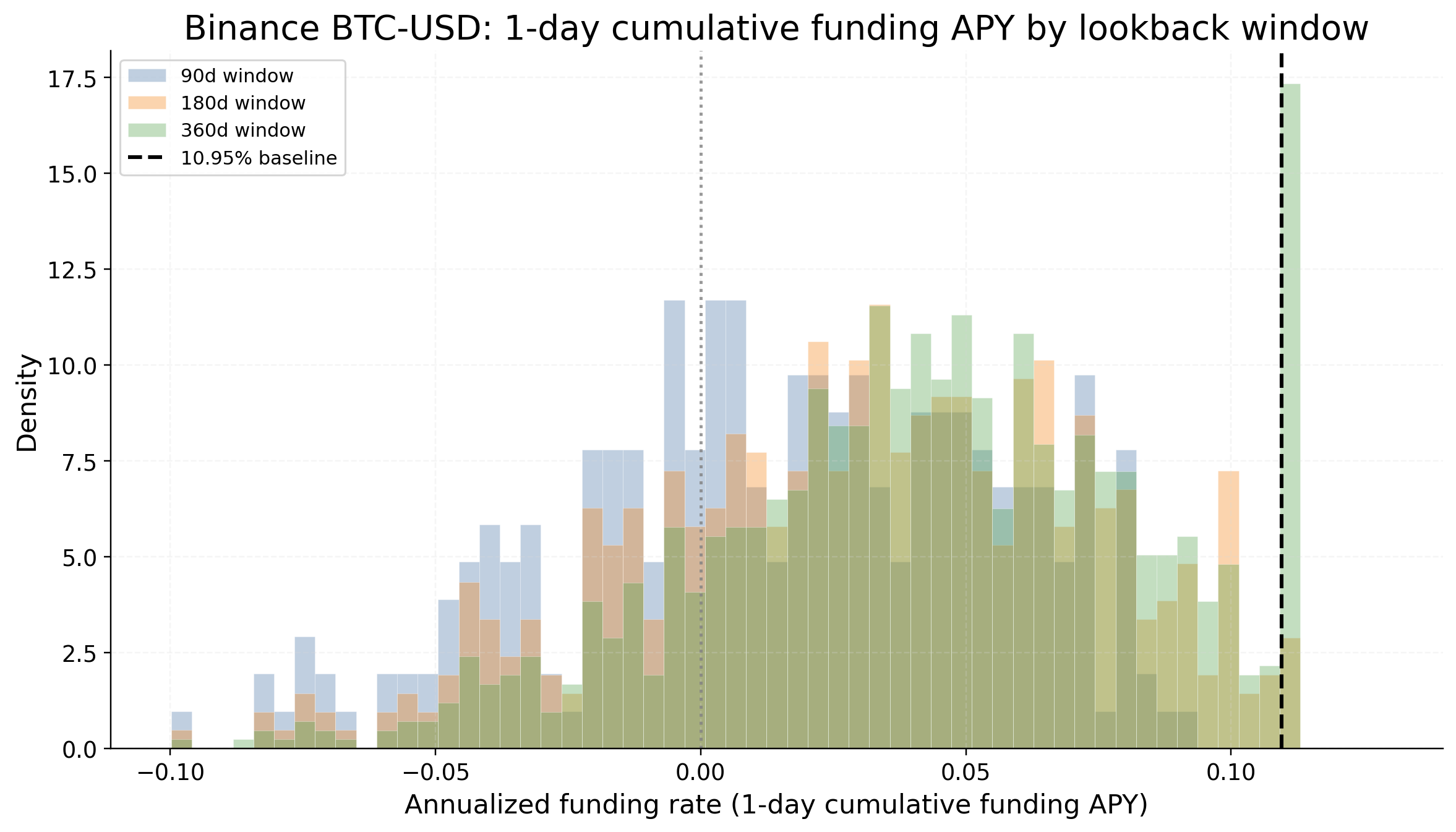}
        \caption{Binance BTC funding 1D APY.}
        \label{fig:raw-funding-binance-btc}
    \end{subfigure}\hspace{0.01\textwidth}%
    \begin{subfigure}[t]{0.495\textwidth}
        \centering
        \includegraphics[width=\linewidth]{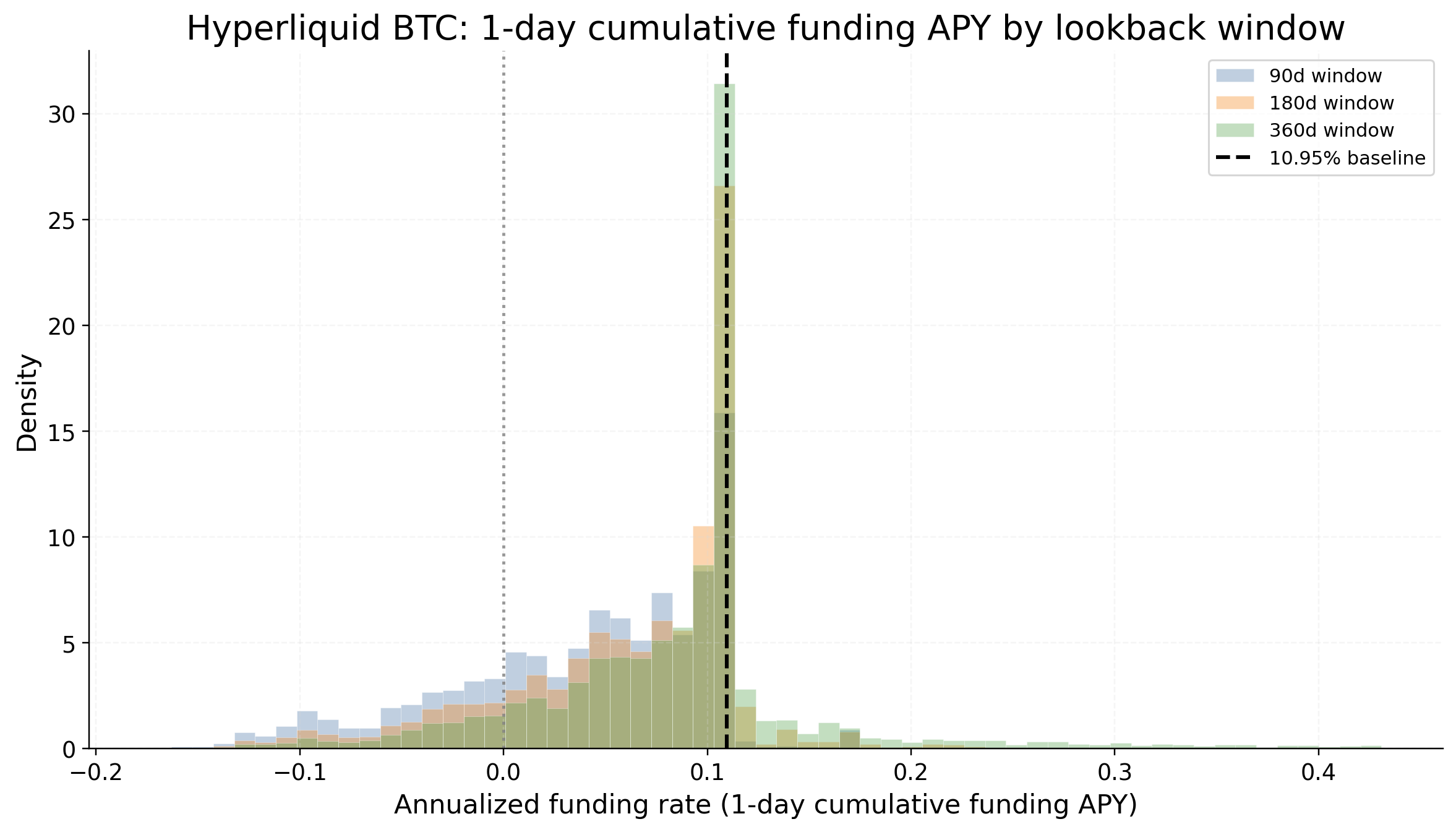}
        \caption{Hyperliquid BTC funding 1D APY.}
        \label{fig:raw-funding-hl-btc}
    \end{subfigure}
    \caption{Raw funding diagnostics for BTC across calibration windows on the centralized and decentralized benchmarks.}
    \label{fig:raw-funding-btc}
\end{figure}

Figure~\ref{fig:raw-funding-btc} reports the raw BTC funding-print distributions in the 90-day, 180-day, and 360-day windows.\vspace{-0.4em}

\section{Historical Calibration and Backtest Inputs}

Tables~\ref{tab:hist-static-v1}--\ref{tab:hist-mc-upper} report the historical calibration objects used by the backtest layer. All tables in this appendix use Binance 2023-01-01 to 2024-12-31 market data together with the current Hyperliquid margin rule $\theta_F=1/(2L_{\max})$.

\begin{table}[H]
\centering
\caption{Historical calibration, Panel A. Variant~1 static grid, Binance 2023-01-01 to 2024-12-31 data, Hyperliquid current margin rule. Values are $\alpha^\star$ across lookback windows and volatility multipliers.}
\label{tab:hist-static-v1}
\small
\begin{tabular}{llccccc}
\toprule
Asset & Lookback & $L_{\max}$ & $\theta_F$ & $1.0\times$ & $1.5\times$ & $2.0\times$ \\
\midrule
BTC  & 30d  & 40 & 0.0125 & 0.152 & 0.209 & 0.262 \\
BTC  & 180d & 40 & 0.0125 & 0.156 & 0.215 & 0.269 \\
ETH  & 30d  & 25 & 0.0200 & 0.161 & 0.219 & 0.272 \\
ETH  & 180d & 25 & 0.0200 & 0.163 & 0.223 & 0.278 \\
LINK & 30d  & 10 & 0.0500 & 0.248 & 0.327 & 0.397 \\
LINK & 180d & 10 & 0.0500 & 0.251 & 0.331 & 0.402 \\
DOGE & 30d  & 10 & 0.0500 & 0.262 & 0.345 & 0.417 \\
DOGE & 180d & 10 & 0.0500 & 0.267 & 0.351 & 0.425 \\
\bottomrule
\end{tabular}
\end{table}

\begin{table}[H]
\centering
\caption{Historical calibration, Panel B. Variant~2 static grid, Binance 2023-01-01 to 2024-12-31 data, Hyperliquid current margin rule. Values are $\alpha_\varepsilon^\star$ across lookback windows and volatility multipliers.}
\label{tab:hist-static-v2}
\small
\begin{tabular}{llccccc}
\toprule
Asset & Lookback & $L_{\max}$ & $\theta_F$ & $1.0\times$ & $1.5\times$ & $2.0\times$ \\
\midrule
BTC  & 30d  & 40 & 0.0125 & 0.124 & 0.173 & 0.220 \\
BTC  & 180d & 40 & 0.0125 & 0.124 & 0.173 & 0.220 \\
ETH  & 30d  & 25 & 0.0200 & 0.132 & 0.183 & 0.230 \\
ETH  & 180d & 25 & 0.0200 & 0.132 & 0.183 & 0.230 \\
LINK & 30d  & 10 & 0.0500 & 0.211 & 0.283 & 0.348 \\
LINK & 180d & 10 & 0.0500 & 0.211 & 0.283 & 0.348 \\
DOGE & 30d  & 10 & 0.0500 & 0.224 & 0.299 & 0.365 \\
DOGE & 180d & 10 & 0.0500 & 0.224 & 0.299 & 0.365 \\
\bottomrule
\end{tabular}
\end{table}

\begin{table}[H]
\centering
\caption{Historical Monte Carlo lower boundary for the backtest layer. Binance 2023-01-01 to 2024-12-31 data, Hyperliquid current margin rule, $h_{\mathrm{liq}}=3$ hours, and $\varepsilon_{\mathrm{liq}}=10^{-4}$.}
\label{tab:hist-mc-lower}
\small
\begin{tabular}{lcccc}
\toprule
Asset & $\alpha^{\dagger}$ & $\alpha_L$ & $\alpha^{\dagger}-\alpha_L$ \\
\midrule
BTC  & 0.173 & 0.045 & 0.128 \\
ETH  & 0.183 & 0.053 & 0.129 \\
LINK & 0.283 & 0.098 & 0.184 \\
DOGE & 0.299 & 0.102 & 0.196 \\
\bottomrule
\end{tabular}
\end{table}

\begin{table}[H]
\centering
\caption{Historical Monte Carlo upper-bound summary. Binance 2023-01-01 to 2024-12-31 data, Hyperliquid current margin rule, 14-day funding horizon.}
\label{tab:hist-mc-upper}
\small
\begin{tabular}{lcccc}
\toprule
Asset & Cost (bps) & $\alpha_U$ & Hit rate & Median hit (d) \\
\midrule
BTC  & 5  & 0.338 & 0.537 & 16.2 \\
BTC  & 10 & 0.502 & 0.187 & 35.8 \\
BTC  & 20 & 0.831 & 0.000 & -- \\
ETH  & 5  & 0.320 & 0.668 & 12.8 \\
ETH  & 10 & 0.458 & 0.343 & 28.9 \\
ETH  & 20 & 0.734 & 0.008 & 51.0 \\
LINK & 5  & 0.416 & 0.792 & 7.6 \\
LINK & 10 & 0.550 & 0.538 & 20.9 \\
LINK & 20 & 0.817 & 0.061 & 48.4 \\
DOGE & 5  & 0.429 & 0.812 & 6.9 \\
DOGE & 10 & 0.559 & 0.577 & 19.8 \\
DOGE & 20 & 0.820 & 0.085 & 46.1 \\
\bottomrule
\end{tabular}
\end{table}

\begin{table}[H]
\centering
\caption{Empirical calibration layers used in the paper.}
\label{tab:calibration-layers-appendix}
\small
\setlength{\tabcolsep}{4pt}
\begin{tabular}{p{0.18\linewidth}p{0.16\linewidth}p{0.24\linewidth}p{0.18\linewidth}p{0.16\linewidth}}
\toprule
Layer & Dates & Data sources & Primary calibrated objects & Role in paper \\
\midrule
Refreshed article calibration & 2025--2026 & Binance and Hyperliquid market data; current venue rules & benchmark $\alpha^{\dagger}$, comparative venue results, dynamic benchmark inputs & descriptive calibration for current comparative statics \\
Historical backtest calibration & 2023--2024 calibration; 2024--2025 validation & Binance spot, mark, and funding data for calibration and baseline backtest; Hyperliquid target margin rule; Hyperliquid funding stream used only in the Section~IX.B counterfactual experiment & frozen ex ante control parameters for historical validation & backtest-layer design and out-of-time validation \\
Live execution calibration & 2025-04-01 to 2025-12-01 & 1inch routed/RFQ spot execution and Hyperliquid live hedging sample & minimum trade size, side-specific execution costs, execution buffer & implementation and execution validation \\
\bottomrule
\end{tabular}
\end{table}

\paragraph*{Additional execution calibration.}
For completeness, Table~\ref{tab:buffer-summary-appendix} reports a compact appendix summary of the additional execution buffer required to reach selected capital-weighted success thresholds in the live sample.

\begin{table}[H]
\centering
\caption{Compact execution buffer summary for the live validation sample. Entries report the minimum additional buffer needed to reach the stated success level.}
\label{tab:buffer-summary-appendix}
\small
\setlength{\tabcolsep}{4pt}
\begin{tabular}{lccc}
\toprule
Target success & All & Buy basis & Sell basis \\
\midrule
90\% capital-weighted success & 7 bps & 6 bps & 8 bps \\
95\% capital-weighted success & 10 bps & 11 bps & 10 bps \\
98\% capital-weighted success & -- & -- & 15 bps \\
\bottomrule
\end{tabular}
\end{table}

\clearpage
\section{Funding-Environment Backtest Summaries}
\label{app:funding-backtest-summaries}

This appendix reports the full ticker-level historical-validation summaries used in Section~IX. The calibrated control rule, historical target parameters, and Hyperliquid margin architecture are held fixed; only the funding input is changed between the Binance-funding and Hyperliquid-funding runs.

\begin{figure}[H]
    \centering
    \includegraphics[width=\linewidth]{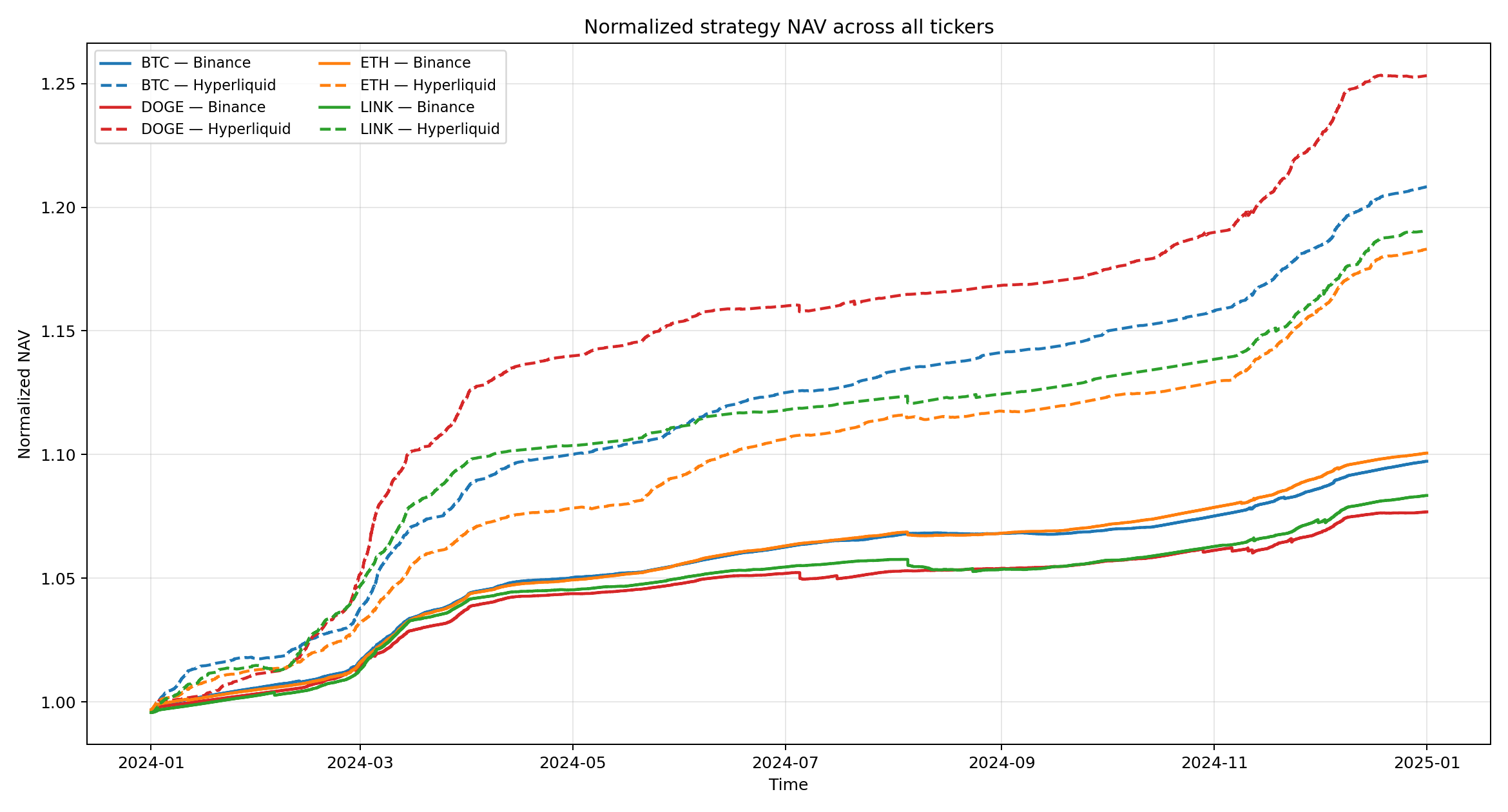}
    \caption{Normalized strategy NAV across all tickers under the two funding environments. Solid lines denote Binance funding; dashed lines denote Hyperliquid funding.}
    \label{fig:all-tickers-nav-appendix}
\end{figure}

\begin{table*}[H]
\centering
\caption{Full ticker-level historical backtest summary.}
\label{tab:all-tickers-summary-full}
\small
\setlength{\tabcolsep}{5pt}
\begin{tabular}{llcccccccc}
\toprule
Ticker & Funding environment & Acc. return (\%) & APY (\%) & Max DD (\%) & Funding APY (\%) & Rebalances & Turnover notional & Avg alpha & Avg leverage \\
\midrule
BTC  & Binance funding      & 9.73 & 9.70 & -0.06 & 10.33 & 5 & 678{,}588 & 0.209 & 4.41 \\
BTC  & Hyperliquid funding  & 20.83 & 20.77 & -0.07 & 21.39 & 5 & 713{,}731 & 0.238 & 3.89 \\
DOGE & Binance funding      & 7.68 & 7.66 & -0.26 & 9.49 & 10 & 2{,}327{,}752 & 0.382 & 1.85 \\
DOGE & Hyperliquid funding  & 25.32 & 25.24 & -0.24 & 26.98 & 9 & 2{,}246{,}528 & 0.408 & 1.66 \\
ETH  & Binance funding      & 10.06 & 10.03 & -0.14 & 10.83 & 7 & 1{,}117{,}774 & 0.228 & 3.95 \\
ETH  & Hyperliquid funding  & 18.31 & 18.25 & -0.18 & 18.99 & 6 & 1{,}023{,}871 & 0.244 & 3.58 \\
LINK & Binance funding      & 8.28 & 8.26 & -0.17 & 9.58 & 6 & 1{,}403{,}134 & 0.337 & 2.15 \\
LINK & Hyperliquid funding  & 18.95 & 18.89 & -0.16 & 20.30 & 6 & 1{,}509{,}586 & 0.364 & 1.92 \\
\bottomrule
\end{tabular}
\end{table*}

\begin{table*}[H]
\centering
\caption{Full rebalancing summary under alternative funding environments.}
\label{tab:all-tickers-rebalance-full}
\small
\setlength{\tabcolsep}{6pt}
\begin{tabular}{llccc}
\toprule
Ticker & Funding environment & Rebalances & Turnover notional & Avg rebalancing notional \\
\midrule
BTC  & Binance funding      & 5 & 678{,}588 & 135{,}718 \\
BTC  & Hyperliquid funding  & 5 & 713{,}731 & 142{,}746 \\
DOGE & Binance funding      & 10 & 2{,}327{,}752 & 232{,}775 \\
DOGE & Hyperliquid funding  & 9 & 2{,}246{,}528 & 249{,}614 \\
ETH  & Binance funding      & 7 & 1{,}117{,}774 & 159{,}682 \\
ETH  & Hyperliquid funding  & 6 & 1{,}023{,}871 & 170{,}645 \\
LINK & Binance funding      & 6 & 1{,}403{,}134 & 233{,}856 \\
LINK & Hyperliquid funding  & 6 & 1{,}509{,}586 & 251{,}598 \\
\bottomrule
\end{tabular}
\end{table*}

\begin{table*}[H]
\centering
\caption{Wide pivot summary for the historical funding-environment comparison.}
\label{tab:all-tickers-pivot-full}
\small
\setlength{\tabcolsep}{4pt}
\begin{tabular}{lcccccccc}
\toprule
Ticker & APY$_{\mathrm{BN}}$ & APY$_{\mathrm{HL}}$ & Funding APY$_{\mathrm{BN}}$ & Funding APY$_{\mathrm{HL}}$ & Rebal$_{\mathrm{BN}}$ & Rebal$_{\mathrm{HL}}$ & Avg Lev$_{\mathrm{BN}}$ & Avg Lev$_{\mathrm{HL}}$ \\
\midrule
BTC  & 9.70 & 20.77 & 10.33 & 21.39 & 5 & 5 & 4.41 & 3.89 \\
DOGE & 7.66 & 25.24 & 9.49 & 26.98 & 10 & 9 & 1.85 & 1.66 \\
ETH  & 10.03 & 18.25 & 10.83 & 18.99 & 7 & 6 & 3.95 & 3.58 \\
LINK & 8.26 & 18.89 & 9.58 & 20.30 & 6 & 6 & 2.15 & 1.92 \\
\bottomrule
\end{tabular}
\end{table*}

\end{document}